\documentstyle[epsfig]{aipproc}

\def\prop{\sim}

\def\lsim{\,\lower.25ex\hbox{$\scriptstyle\sim$}\kern-1.30ex%
\raise 0.55ex\hbox{$\scriptstyle <$}\,}
\newcommand {\pom} {I\!\!P}
\newcommand {\pomsub} {{\scriptscriptstyle \pom}}

\newcommand {\apom} {\alpha_{\pomsub}}

\newcommand {\aprime} {\alpha^\prime_\pomsub}

\newcommand{\rfour}{\mbox{$r^{04}_{00}$}}

\newcommand{\rfivecomb}{\mbox{$r^5_{00} + 2 r^5_{11}$}}
\newcommand{\ronecomb}{\mbox{$r^1_{00} + 2 r^1_{11}$}}
\newcommand{\tprim}{\mbox{$t^\prime$}}
\newcommand{\tlc}{\mbox{$t$}}
\newcommand\units[1]{\,\mathrm{#1}} 
\newcommand\fig[1]{Fig.\,\ref{fig:#1}} 

\newcommand\qq[1]{equation (\ref{qq:#1})}

\def\ap#1#2#3   {{\em Ann. Phys. (NY)} {\bf#1} (#2) #3.}   
\def\apj#1#2#3  {{\em Astrophys. J.} {\bf#1} (#2) #3.} 
\def\apjl#1#2#3 {{\em Astrophys. J. Lett.} {\bf#1} (#2) #3.}
\def\app#1#2#3  {{\em Acta. Phys. Pol.} {\bf#1} (#2) #3.}
\def\ar#1#2#3   {{\em Ann. Rev. Nucl. Part. Sci.} {\bf#1} (#2) #3.}
\def\cpc#1#2#3  {{\em Computer Phys. Comm.} {\bf#1} (#2) #3.}
\def\epj#1#2#3  {{\em Eur. Phys. J.} {\bf#1} (#2) #3}
\def\err#1#2#3  {{\it Erratum} {\bf#1} (#2) #3.}
\def\ib#1#2#3   {{\it ibid.} {\bf#1} (#2) #3.}
\def\jmp#1#2#3  {{\em J. Math. Phys.} {\bf#1} (#2) #3.}
\def\ijmp#1#2#3 {{\em Int. J. Mod. Phys.} {\bf#1} (#2) #3}
\def\jetp#1#2#3 {{\em JETP Lett.} {\bf#1} (#2) #3.}
\def\jpg#1#2#3  {{\em J. Phys. G.} {\bf#1} (#2) #3.}
\def\mpl#1#2#3  {{\em Mod. Phys. Lett.} {\bf#1} (#2) #3.}
\def\nat#1#2#3  {{\em Nature (London)} {\bf#1} (#2) #3.}
\def\nc#1#2#3   {{\em Nuovo Cim.} {\bf#1} (#2) #3.}
\def\nim#1#2#3  {{\em Nucl. Instr. Meth.} {\bf#1} (#2) #3}
\def\np#1#2#3   {{\em Nucl. Phys.} {\bf#1} (#2) #3}
\def\pcps#1#2#3 {{\em Proc. Cam. Phil. Soc.} {\bf#1} (#2) #3.}
\def\pl#1#2#3   {{\em Phys. Lett.} {\bf#1} (#2) #3}
\def\prep#1#2#3 {{\em Phys. Rep.} {\bf#1} (#2) #3.}
\def\prev#1#2#3 {{\em Phys. Rev.} {\bf#1} (#2) #3}
\def\prl#1#2#3  {{\em Phys. Rev. Lett.} {\bf#1} (#2) #3}
\def\prs#1#2#3  {{\em Proc. Roy. Soc.} {\bf#1} (#2) #3.}
\def\ptp#1#2#3  {{\em Prog. Th. Phys.} {\bf#1} (#2) #3.}
\def\ps#1#2#3   {{\em Physica Scripta} {\bf#1} (#2) #3.}
\def\rmp#1#2#3  {{\em Rev. Mod. Phys.} {\bf#1} (#2) #3}
\def\rpp#1#2#3  {{\em Rep. Prog. Phys.} {\bf#1} (#2) #3.}
\def\sjnp#1#2#3 {{\em Sov. J. Nucl. Phys.} {\bf#1} (#2) #3}
\def\spj#1#2#3  {{\em Sov. Phys. JEPT} {\bf#1} (#2) #3}
\def\spu#1#2#3  {{\em Sov. Phys.-Usp.} {\bf#1} (#2) #3.}
\def\zp#1#2#3   {{\em Zeit. Phys.} {\bf#1} (#2) #3}

\begin{document}
\title{ Exclusive Vector Meson Production at HERA}
\author{Arik Kreisel\thanks{Email: arikk@alzt.tau.ac.il} \thanks{
LISHEP 2002 - Session C:Workshop On Diffractive Physics - February 4 - 8 2002
 Rio de Janeiro - RJ - Brazil}\\ for the H1 and the ZEUS Collaborations}
\address{School of Physics and Astronomy,\\
Raymond and Beverly Sackler Faculty of Exact Sciences\\
Tel Aviv University, Tel Aviv, Israel.}

\maketitle

\begin{abstract}

An extended study of exclusive vector meson production in $ep$
interactions has been performed by the H1 and the ZEUS collaborations
at the HERA collider. Recent measurements are reported and discussed
within the framework of the dipole model and pQCD.
\end{abstract}

\section{Introduction}

The sharp rise of the electromagnetic proton structure function,
$F_2$, toward low values of Bjorken $x$, discovered at
HERA~\cite{zeusf2,h1f2}, and the observation of a large fraction of
diffractive-like events~\cite{zeusdiff,h1diff} are attributed to a
large gluon density in the proton at very low $x$ values, typically
 $x<0.01$. The rise of $F_2$ with decreasing $x$ can be
accommodated by the QCD, DGLAP evolution equations~\cite{DGLAP} in NLO
down to momentum transfer squared $Q^2 \simeq 1
\units{GeV^2}$~\cite{DGLAPlowq2,CTEQ5M}. This suggests that
perturbative effects set in at relatively low values of the
interaction scale.  However, the validity of the DGLAP evolution
equation is established through fits to data which involve many
unknown parameters and therefore it may be doubtful. Exclusive vector
meson (V) production at high $Q^2$ has been proposed as an alternative
method to infer the gluon content of the proton~\cite{brodsky}.

High energy elastic V production in Deep Inelastic Scattering (DIS)
may be described, in the rest frame of the proton, by the fallowing
 sequence of
 happenings~\cite{strikfurtHA}. The incoming lepton
emits a virtual photon, which subsequently fluctuates into a
$q\bar{q}$ pair. The life time of such a quark pair fluctuation is
long enough so that it is the pair that elastically scatters off the
proton and evolves, long after the interaction, into a V state.

The character of the interaction of the $q\bar{q}$ pair with the
proton depends on the transverse momenta of the pair. If the
transverse momentum is large, the spatial transverse separation
between the quarks is small and it forms a color dipole, whose
interaction with the proton may be calculated
perturbatively~\cite{strikfurt-dipole}. The leading process is two
gluon exchange. If the transverse momentum is small, the color dipole
is large and perturbative calculations do not apply. In this case the
interaction looks similar to hadron-hadron elastic scattering and the
process should proceed through Pomeron exchange as expected from Regge
phenomenology~\cite{collins}. 

The $q\bar{q}$ wave function of the virtual photon 
 depends on the polarization of the virtual
photon. For longitudinally polarized photons, small transverse size
$q\bar{q}$ dominate. The opposite is true for transversely polarized
photons. The
attractive features of elastic V production is that, at high $Q^2$,
the longitudinal component of the virtual photon dominates. The
interaction cross section for the latter can be, in principle, fully
calculated in perturbative QCD. Moreover, for heavy vector mesons,
like the $J/\Psi$ or the $\Upsilon$, perturbative calculations apply
even at $Q^2=0$, as the smallness of the dipole is guaranteed by the
mass of the quarks.

Independently of particular calculations~\cite{reviewVM,ivanov}, in the
region dominated by perturbative QCD, the following features are
predicted:
\begin{itemize}
\item the total $\gamma^\star p \rightarrow V p$ cross section,
  $\sigma_{\gamma^\star p \rightarrow V p}$, exhibits a steep rise
  with $W$, the photon-proton center-of-mass energy, that can be
  parameterized as $\sigma \sim W^{\delta}$, with $\delta$ increasing
  with $Q^2$;
\item the $Q^2$ dependence, which for a longitudinally polarized
  photon is expected to behave like $Q^{-6}$, becomes milder, more
  like $Q^{-4}$, due to the sharp increase of the gluon density with
  $Q^2$;
\item the distribution of $t$, the four momentum transfer squared at
  the proton vertex, becomes universal, with little or no dependence
  on $W$ or $Q^2$;
\end{itemize}

All these features have been observed at 
HERA~\cite{review-data,osaka,h1apom,z-disrho} in
the exclusive production of $\rho^0$, $\phi$, $\omega$ and $J/\Psi$
mesons. However, the agreement between measurements and theory is far
from perfect.  There are many factors that may spoil the agreement.
First and foremost, the measured cross sections contain the possibly
soft transverse component of the virtual photon, which has to be
modeled. In addition, the perturbative calculations have the
following uncertainties:
\begin{itemize}
\item the calculation of $\sigma_{\gamma^\star p \rightarrow Vp}$
  involves the so-called skewed parton distributions~\cite{skewed},
  which are not yet well tested and involve gluon distributions
  outside the range which is constrained by global parton density
  analyses~\cite{jhep:103:45};
\item higher order corrections have not been fully calculated,
  therefore the overall normalization is uncertain and the scale at
  which the gluons are probed is not known;
\item the fast rise of $\sigma_{\gamma^\star p \rightarrow V p}$
  implies the presence of the real part of the scattering amplitude,
  which is not fully known;
\item the wave functions of the vector mesons are not fully under
  control.
\end{itemize}
In spite of all these problems, it is generally felt that precise
measurements of $\sigma_{\gamma^\star p \rightarrow V p}$ as a
function of $W$, $Q^2$, $t$ and the mass of the vector meson, 
$M_V$, with a
separation into longitudinal and transverse components~\cite{amirim},
will help to resolve the theoretical uncertainties and ultimately lead
to a better understanding of the parton distributions in the proton as
well as of the dynamics of high energy interactions in the presence of
a large scale.

One of the challenges in confronting perturbative QCD calculations
with data is the ability to establish a region where hard interactions
dominate over the soft component. The soft component is believed to be
well described by Regge phenomenology, according to which at high
energy, the Pomeron exchange dominates in the production of
diffractive states. The parameters of the Pomeron trajectory are known
from measurements of total cross sections in hadron-hadron
interactions and elastic proton-proton measurements.  It is usually
assumed that the Pomeron trajectory is linear in $t$ and has the
following form:
\begin{equation}
\apom (t) = \apom(0) + \aprime t \, .
\end{equation}
The parameter $\apom(0)$ determines the energy behavior of the
total cross section,
\begin{equation}
\sigma_{\rm tot} \prop s^{\apom(0)-1} \, ,
\end{equation}
and $\aprime$ describes the increase of the exponential slope $b$ of
the $t$ distribution with increasing $s$. The value of $\aprime$ is
expected to be
inversely proportional to the typical transverse momenta squared of
the partons participating in the exchanged
trajectory~\cite{Gribov:1961}.  Indeed a large value of $\aprime$ 
would suggest
the presence of low transverse momenta, typical of soft interactions.
The latest fits of $\apom(0)$~\cite{cudell} and $
\aprime$~\cite{landshoff-aprime} are give:
\begin{eqnarray}
\apom(0) &=& 1.096 \pm 0.003 \\
\aprime &=& 0.25 \units{GeV^{-2}} \, .
\end{eqnarray}
These values for the trajectory parameters describe the so called soft
Pomeron trajectory.  The non-universality of $\apom(0)$ has been
established in DIS, where the slope of the rise of the $\gamma^\star
p$ total cross section with $W$ has a pronounced $Q^2$
dependence~\cite{h1lambda,zeuslambda}. The issue of the $\aprime$
dependence on the hardness of the interaction may be addressed in the
study of exclusive V production at HERA. The value of $\aprime$ may be
determined from the $W$ dependence of $b$, since $b$ is expected to
behave as
\begin{equation}
b(W) = b_0 +4 \aprime \ln\frac{W}{W_0}  \, .
\end{equation}
The parameter $\aprime$ may also be derived from the $W$ dependence of the
differential  cross section $d\sigma /dt$ at fixed $t$,
\begin{equation}
\frac{d\sigma}{dt}(W) = F(t) W^{2[2\apom(t)-2]} \, ,
\end{equation}
where $F(t)$ is a function of $t$ only.  The latter approach has the
advantage that no assumption needs to be made about the $t$
dependence. The measurements of $\apom(t)$ in exclusive $J/\psi$
photoproduction show that, while $\apom(0)$ for $J/\psi$ is larger the
soft $\apom(0)$ value, the value of $\aprime$ for $J/\psi$ is
smaller~\cite{h1apom,levyapom,zeusapom} than the soft value.

Another way to investigate the contribution of hard and soft
components in V production is by projecting out the interactions
induced by the longitudinally and transversely polarized
virtual photons. Due to the respective wave functions structure, small
$q\bar{q}$ configurations are dominantly longitudinal while large
configurations are dominantly transverse.  If one assumes that
exclusive V production proceeds through $s$-channel helicity conservation
(SCHC), the separation into the longitudinal and transverse components
is possible. At small $t$, the hypothesis of SCHC has been directly
tested in the data, and small deviations have been
observed~\cite{H199-010,zeusschc}. However, the deviations are small
enough not to jeopardize the decomposition of the cross sections into
the longitudinal and transverse components. 

The angular distributions of the decay
products of the vector mesons give access to the spin density matrix
elements, which are bilinear combinations of the helicity amplitudes 
$T_{\lambda_V \lambda_{\gamma}}$ where $\lambda_V$ ($\lambda_{\gamma}$)
is the V (virtual photon) helicity~\cite{schilling}.
As $t$ increases, a possible change in helicity is more likely due
to the change in the V's direction and the transfer of the transverse
momentum carried by the gluons. The following features are expected by
pQCD models~\cite{ivanov,royen,niko} in DIS for $|t|\lsim Q^2$: 
\begin{itemize}
\item a constant ratio  of the helicity conserving amplitudes,  
$|T_{11}| \ / \ |T_{00}|$, with $t$ ;
\item a $\sqrt {|t|}$ dependence for the ratio of the single helicity
flip to the non-flip amplitudes
$|T_{01}| \ / \ |T_{00}|$ and $|T_{10}| \ / \ |T_{00}|$;
\item a linear $t$  dependence for the ratio of the double flip to
the non-flip amplitudes
$|T_{1-1}| \ / \ |T_{00}|$;
\item the hierarchy 
\begin{equation}
|T_{00}| > |T_{11}| > |T_{01}| > |T_{10}| > |T_{1-1}| \ . 
                                \label{eq:hierarchy}
\end{equation}
\end{itemize}

The differential cross section, $\frac{{\it d} \sigma}{{\it d}t}$,
of V production has an exponentially falling cross section at low $t$
and a power like behavior at high $t$. An exponential behavior is
associated with a Gaussian charge distribution and the slope, $b$, of
the exponent is related to the width of the distribution and is a
measure of the interaction size. In a naive picture for elastic
scattering, $b$ has contributions from the size of the $q\bar{q}$ pair
which decreases with $Q^2$ and the constant size for the proton. At
high $Q^2$, for all type of elastic V production,  $b$ should reach an
asymptotic value equal to the proton size. 

The cross
section for elastic V photo or electroproduction, with
small transfer momentum to the proton, has the form~\cite{t_Ryskin}
\begin{equation}
\frac{{\it d}\sigma_{\gamma p \rightarrow V p}}{{\it d}t}=
F^2_p(t)F^2_V(t,Q^2)|A(W,t,Q^2)|^2 \, ,
\label{qq:Ryskin1}
\end{equation}
where $F_p(t)$ and $F_V(t)$ are the form factors of the proton and 
the V,
respectively, that account for the probabilities of their elastic
production, and $A(W,t)$ is the amplitude of the constituent
interaction.  The cross section of  proton
dissociation V production does not contain the proton form factor
\begin{equation}
\frac{{\it d}\sigma_{\gamma p \rightarrow V Y}}{{\it d}t}=
F^2_V(t,Q^2)|A(W,t,Q^2)|^2 \, .
\label{qq:Ryskin2}
\end{equation}
Vertex factorization~\cite{factorization} predicts that
the ratio of proton dissociation over elastic differential cross section is
 $Q^2$ independent. The ratio of \qq{Ryskin1} to \qq{Ryskin2} is equal to
the square of the proton form factor.

\section{$W$ dependence of V production}
\par
The dependence of the cross section $\sigma_V=\sigma_{\gamma^{(\star)}
p \rightarrow Vp}$ on the center of mass energy, $W$, is shown in
\fig{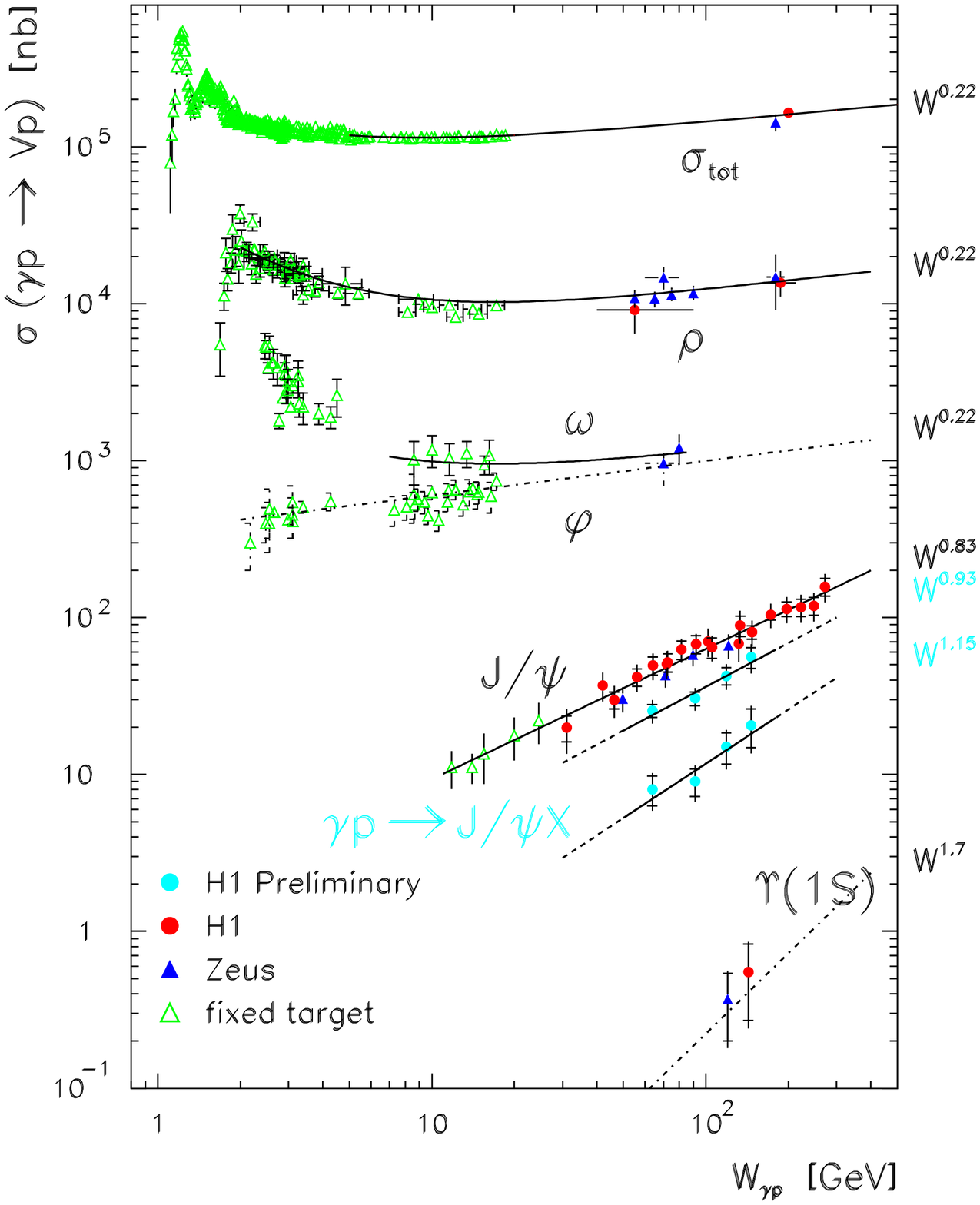} for photoproduction of various 
vector mesons~\cite{allvm}.
While the $W$ dependence of light vector mesons ($\rho$ $\omega$ $\phi$)
 is $\sigma_V(W) \propto W^{0.22}$, as
expected from Regge phenomenology, the $J/\psi$ cross section has a
steep rise with $W$, which is a signature of a hard process, as
predicted by pQCD.

\begin{figure}[hbt]
\begin{center}
\epsfxsize=0.3\vsize
\epsfbox{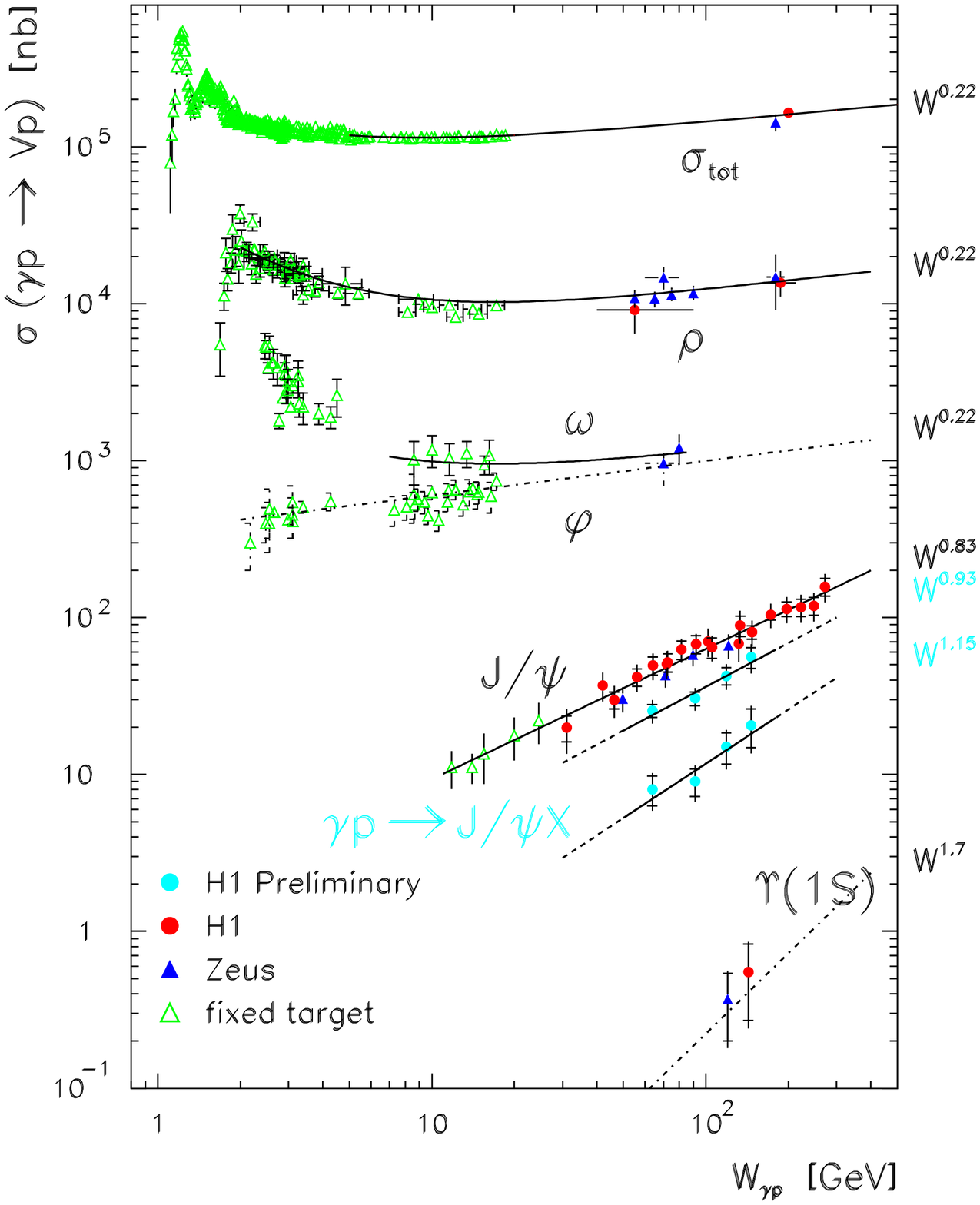}
\caption{The photoproduction cross section as a function
  of $W$ , for different vector mesons. The lines show a $W$
  dependence, with $\delta$ values as indicated.}
\label{fig:w_php_allvm.ps}
\end{center}
\end{figure}

A change in the $W$ dependence is also seen when moving from low to
high $Q^2$, as shown in \fig{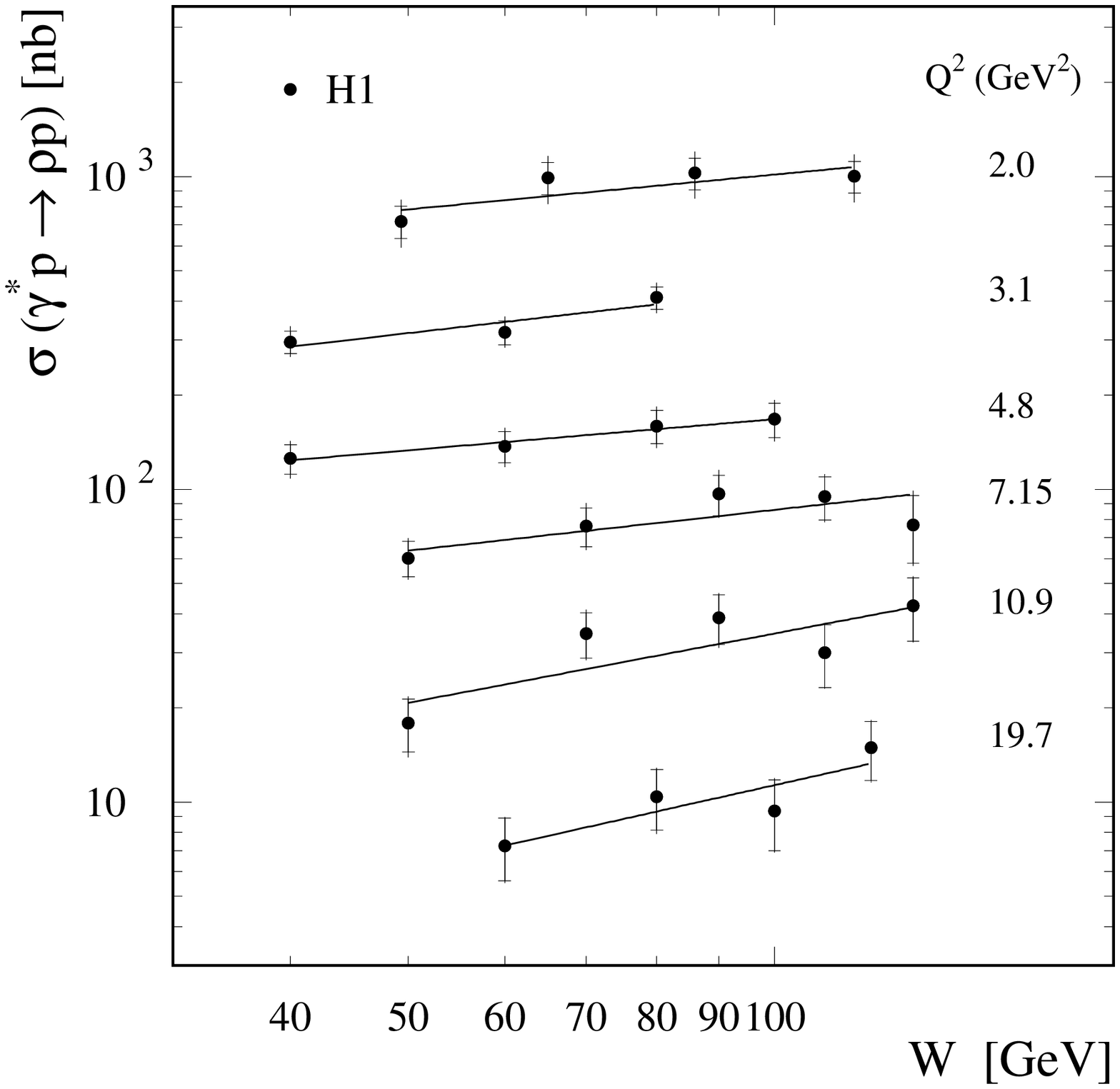}, where the $W$
dependence of $\sigma_\rho$ for various $Q^2$ values~\cite{H199-010}
is plotted .  For each $Q^2$ value the $W$ dependence of $\sigma_\rho$
is fitted with a form $W^{\delta}$.  The parameter $\delta$ is related
to the exchanged trajectory $\apom(t)$ by $\delta=4(\alpha(<t>)-1)$.
To extract $\apom(0)$ from $\delta$, $<|t|>=1/b$ is taken from
measured values~\cite{H199-010} and the value $\aprime=0.25
\units{GeV^{-2}}$ is assigned. The resulting $\apom(0)$ is plotted as a
function of $Q^2$ in \fig{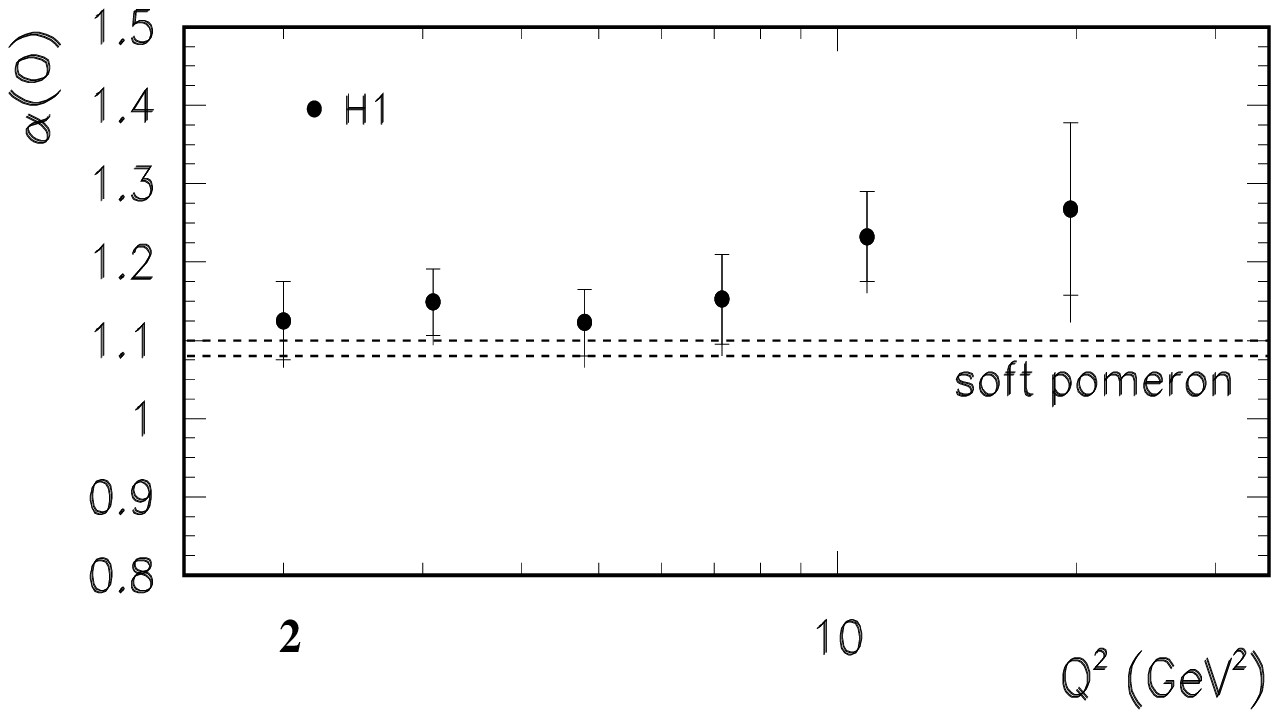} and a marked increase of
$\apom(0)$ with $Q^2$ is observed.

\begin{figure}[hbt]
\vspace*{-5mm}
\begin{minipage}{0.45\hsize}
\epsfxsize=1\hsize
\epsfbox{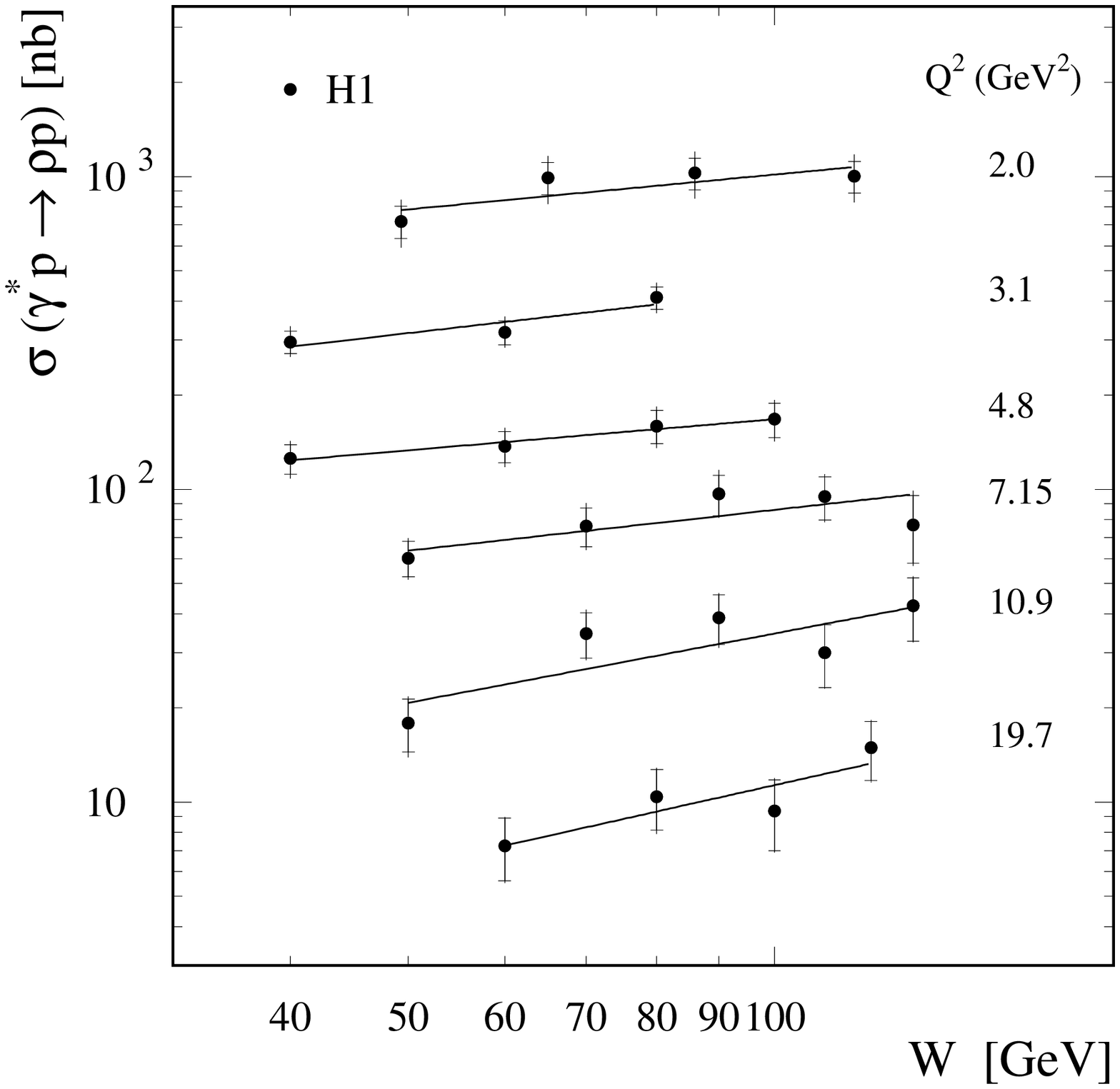}
\caption{The cross section $\sigma_{\gamma^\star p \rightarrow \rho p}$
  as a function of $W$, for several values of $Q^2$. The inner error
  bars are statistical and the full error bars include the systematic
  errors added in quadrature. The lines correspond to a fit of the form
  $\sigma_{\rho} \propto W^{\delta}$.}
\label{fig:h199-010f23.eps}
\end{minipage}
\hspace*{2mm}
\begin{minipage}{0.45\hsize}
\vspace*{-6mm}
\epsfxsize=1\hsize
\epsfbox{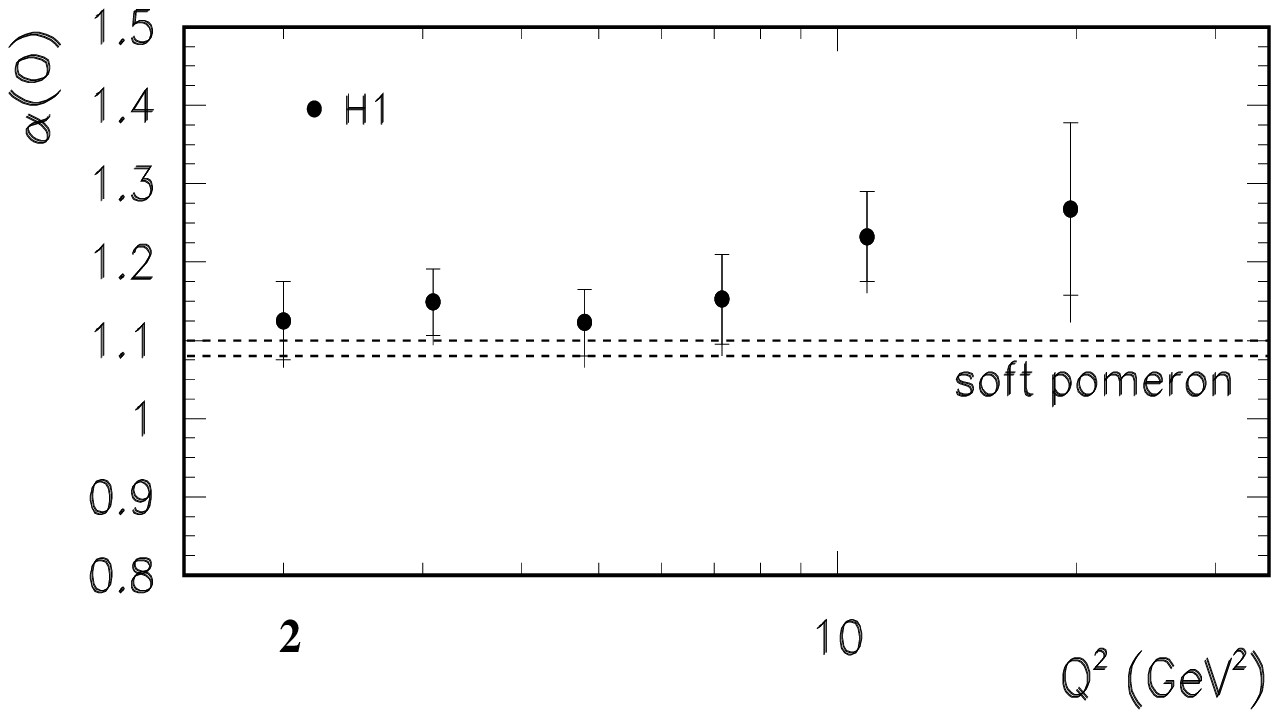}
\caption{The $Q^2$ dependence of the intercept $\apom(0)$. 
  The inner error bars are the statistical and non-correlated
  systematic uncertainties and the outer error bares include the
  variation of the intercept $\apom(0)$ when assuming $\aprime=0$,
  added in quadrature. The dashed lines represent the range of values
  obtained for the ``soft Pomeron''
  intercept~\protect\cite{landshoff-aprime}.}
\label{fig:h199-010f24.eps}
\end{minipage}
\end{figure}

\begin{figure}[htb]
\vspace*{-5mm}
\begin{minipage}{0.45\hsize}
\epsfxsize=1\hsize 
\epsfbox{cross_section_theory.eps}
\caption{The exclusive $J/\psi$ photoproduction cross section as a function
  of $W$.  The inner bars indicate the statistical uncertainties, the
  outer bars are the statistical and systematic uncertainties added in
  quadrature. The lines are predictions of pQCD
  calculations.}
\label{fig:cross_section_theory.eps}
\end{minipage}
\hspace*{2mm}
\begin{minipage}{0.45\hsize}
\vspace*{-6mm}
\epsfxsize=1\hsize 
\epsfbox{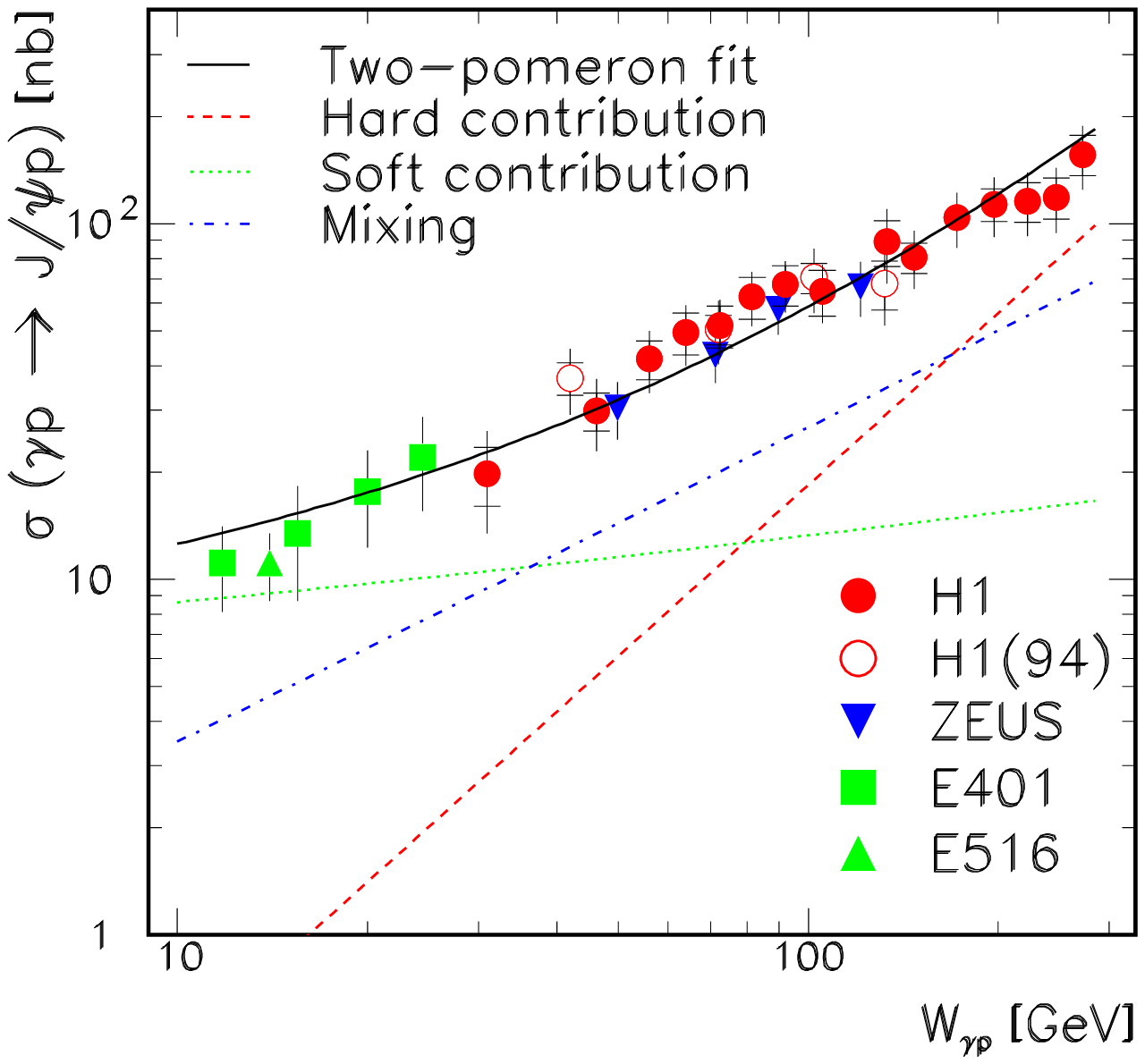}
\caption{The cross section $\sigma(\gamma p \rightarrow J/\psi p)$ 
  versus $W$. The full line is the prediction of the two-Pomeron model
  by Donnachie and Landshoff.  The separate contributions of the hard,
  soft and mixing terms are also indicated.}
\label{fig:w_jpsi_fit1.ps}
\end{minipage}
\end{figure}

The comparison between the measurements of $\sigma_{J/\psi}$ 
in photoproduction and
various pQCD calculations~\cite{jhep:103:45,mrt}, based on different
gluon distributions, is shown in
\fig{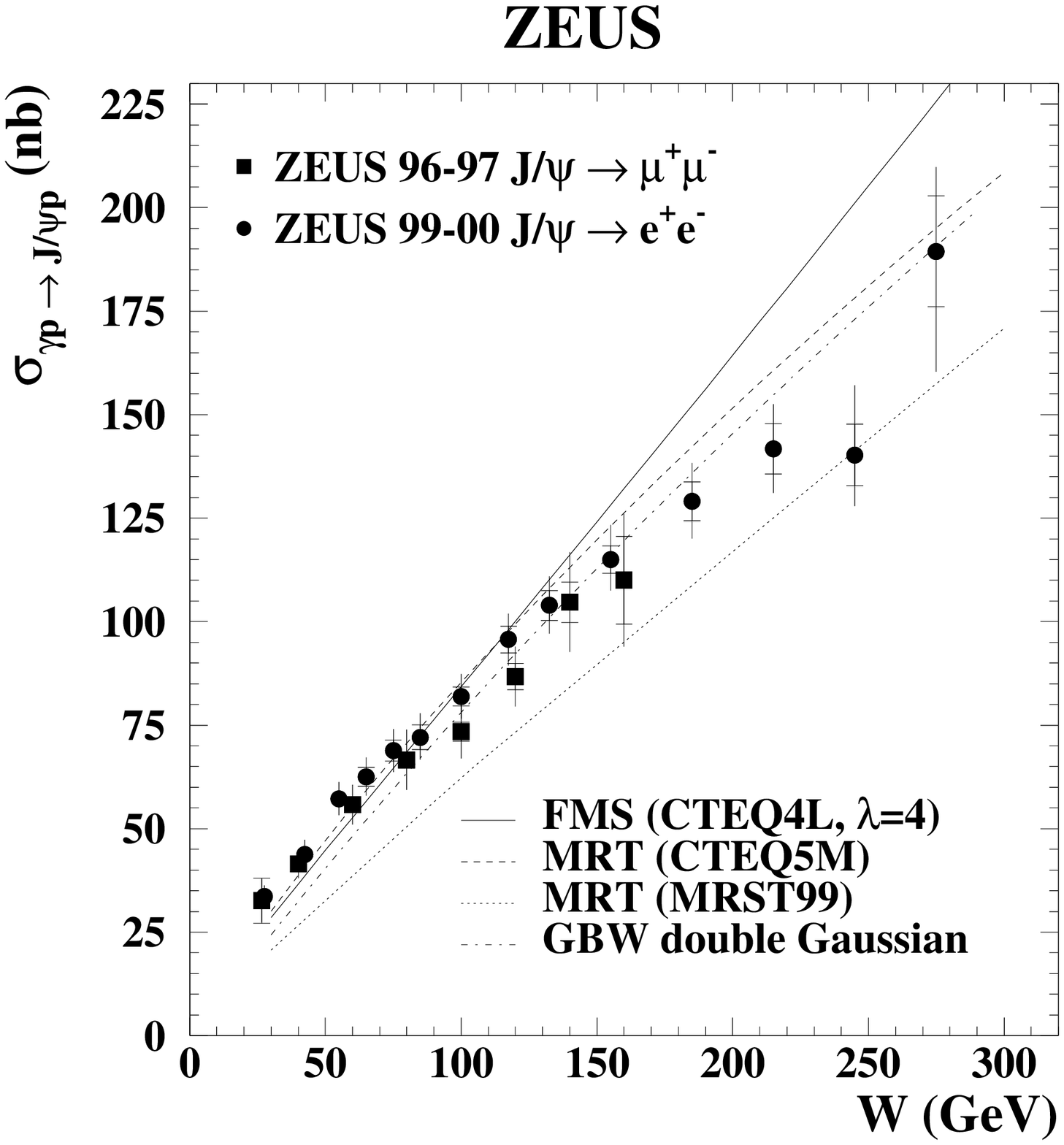}~\cite{ZEUS_J/psi_PHP}.  The pQCD
calculations are known to suffer from large theoretical uncertainties,
therefore no discrimination between the gluon distributions can be
made from this comparison. Also shown in
\fig{cross_section_theory.eps} is the result of a
calculation~\cite{dipole_model_1} based on the Golec-Biernat
W\"{u}sthoff (GBW) model~\cite{dipole_model_2} of the photon, whose
parameters were fitted to the measurements of the inclusive structure
function.  The GBW model gives a good representation of the data,
assuming a double Gaussian wave function of the $J/\psi$.

Although the $W$ dependence of $J/\psi$ photoproduction cannot be
explained by the exchange of a universal soft Pomeron, the
data can be fitted to the two-Pomeron model~\cite{D_and_L} as shown in
\fig{w_jpsi_fit1.ps}~\cite{H1_J/psi_PHP}. The parameters of the
established soft and the proposed hard Pomeron trajectories are
$(\alpha(0),\aprime)_{soft}=(1.08,0.25 \units{GeV^{-2}})$ and
$(\alpha(0),\aprime)_{hard}=(1.418,0.1 \units{GeV^{-2}})$.  The
relative contributions of the hard, soft and mixing terms of the model
are found to vary between $0.1:0.5:0.4$ at $W=30 \units{GeV}$ to
$0.5:0.1:0.4$ at $W=250 \units{GeV}$.

In \fig{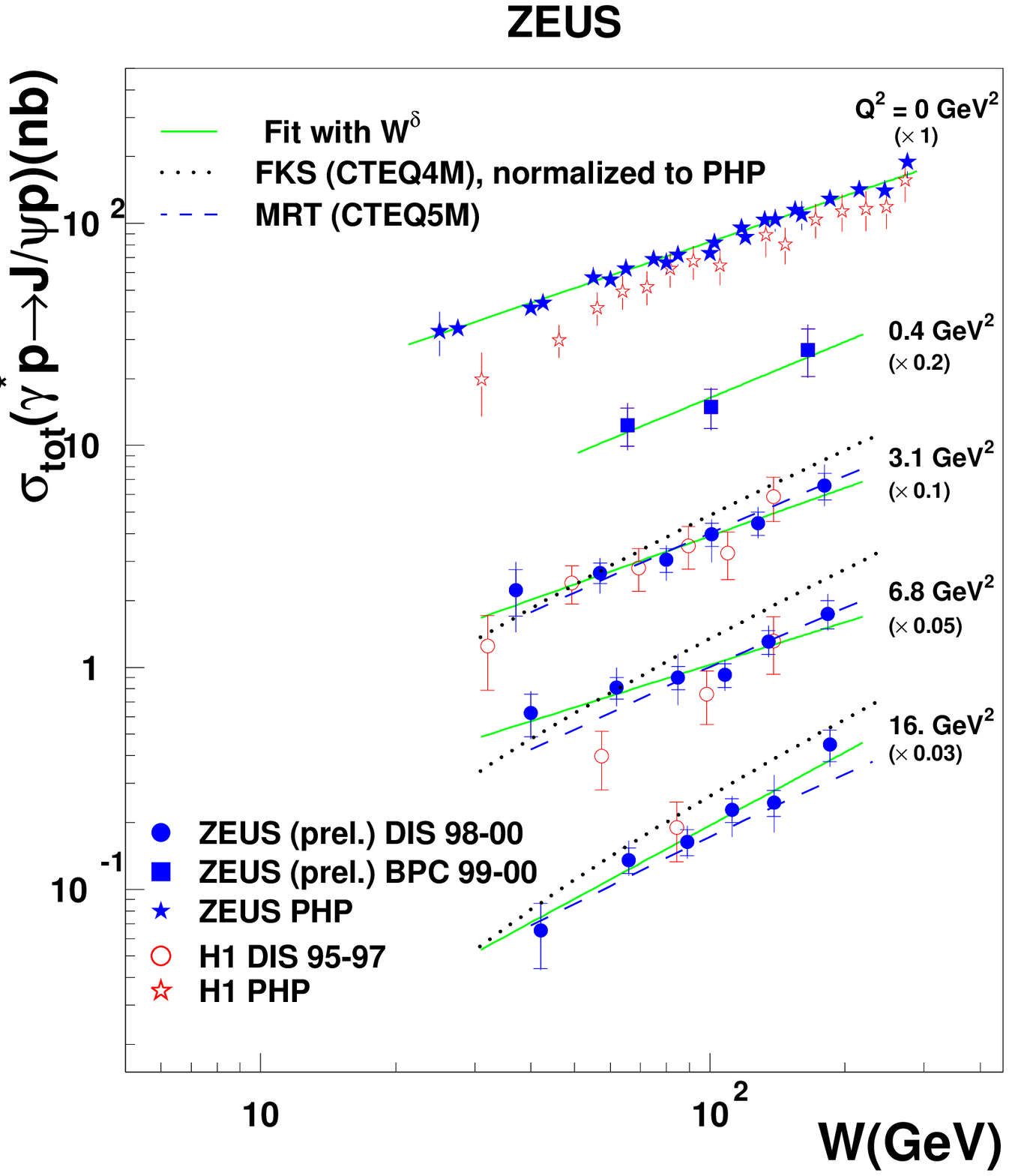} the $W$ dependence of $\sigma_{J/\psi}$ is
shown for various $Q^2$ values~\cite{z-disjpsi}. 
The measurements are compared
with theoretical predictions of Frankfurt et al.  (FKS)~\cite{fks}
using CTEQ4M~\cite{Lai}, and of Martin et al.  (MRT)~\cite{mrt} using
the CTEQ5M~\cite{CTEQ5M} parametrization.  The calculations are consistent
with the data within the uncertainty of the measurements.
Also shown 
in \fig{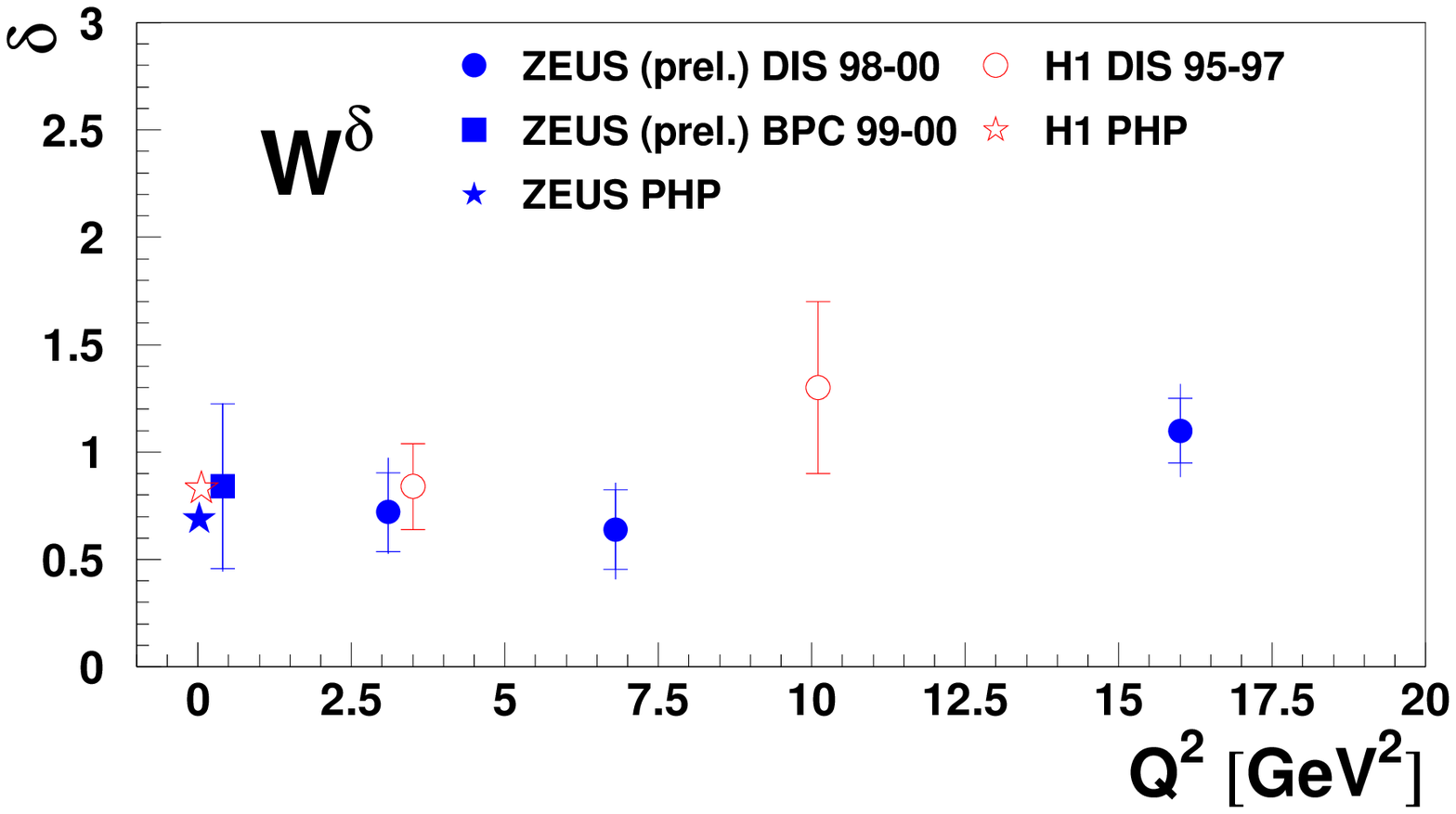} are the
values of $\delta$, obtained by fitting in each $Q^2$, the functional
form $\sigma_{J/\psi} \propto W^{\delta}$.  The results may indicate
a slight increase in $\delta(Q^2)$ at high $Q^2$.

\begin{figure}[hbt]
\vspace*{-5mm}
\begin{minipage}{0.45\hsize}
\epsfxsize=1\hsize
\epsfbox{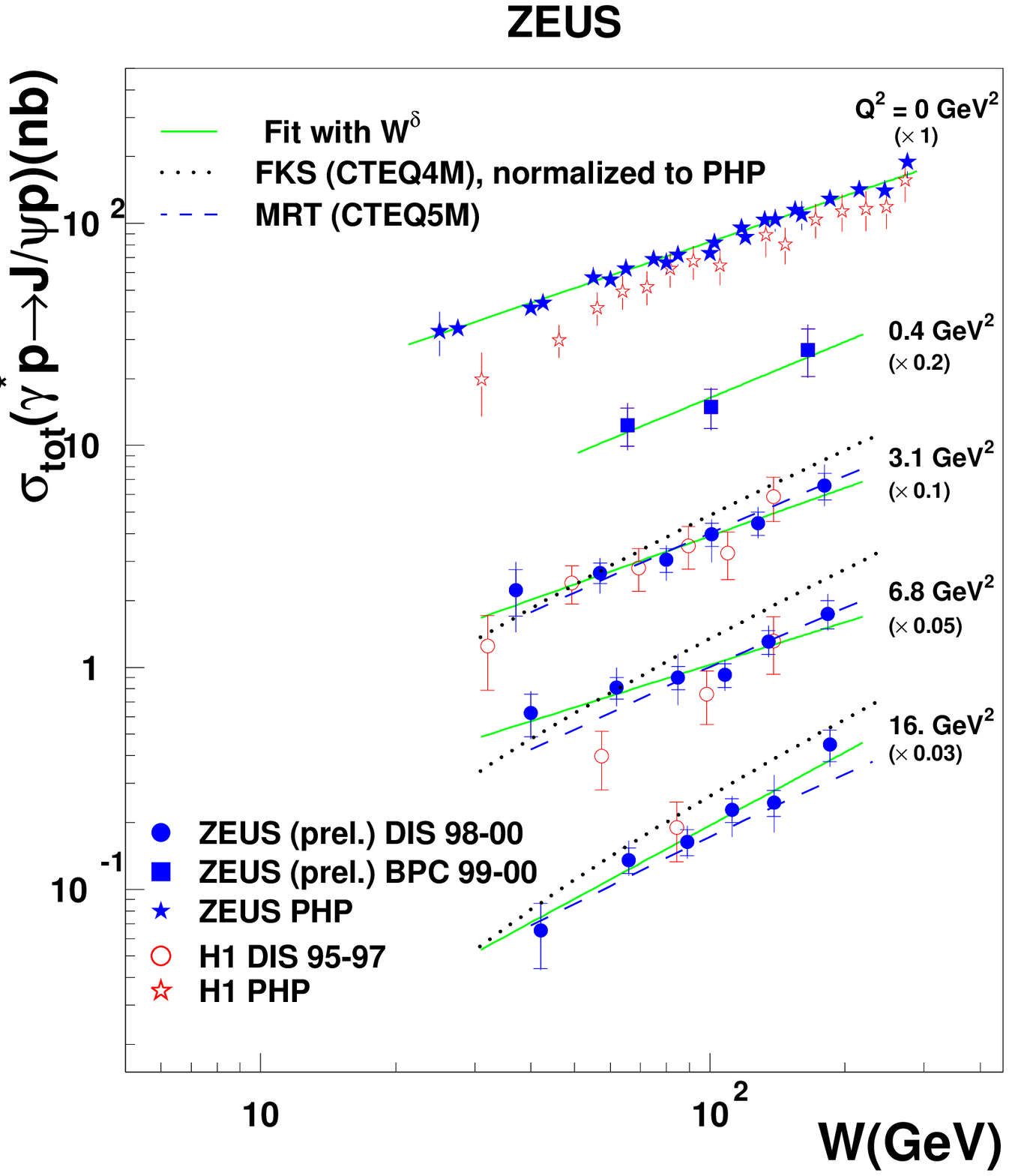}
 \caption{$W$ dependence of the cross section 
  $\sigma(\gamma^\ast p \rightarrow J/\psi p)$ compared with
  theoretical predictions for $Q^2=3.1,\, 6.8$ and 
$16\units{GeV^2}$. The
  solid lines indicate the results of fits to the function
  $W^{\delta}$. The measurements from photoproduction are also shown
  for comparison.}
\label{fig:jpsidis-xsec.eps}
\end{minipage}
\hspace*{2mm}
\begin{minipage}[b]{0.45\hsize}
\vspace*{-6mm}
\epsfxsize=1\hsize
\epsfbox{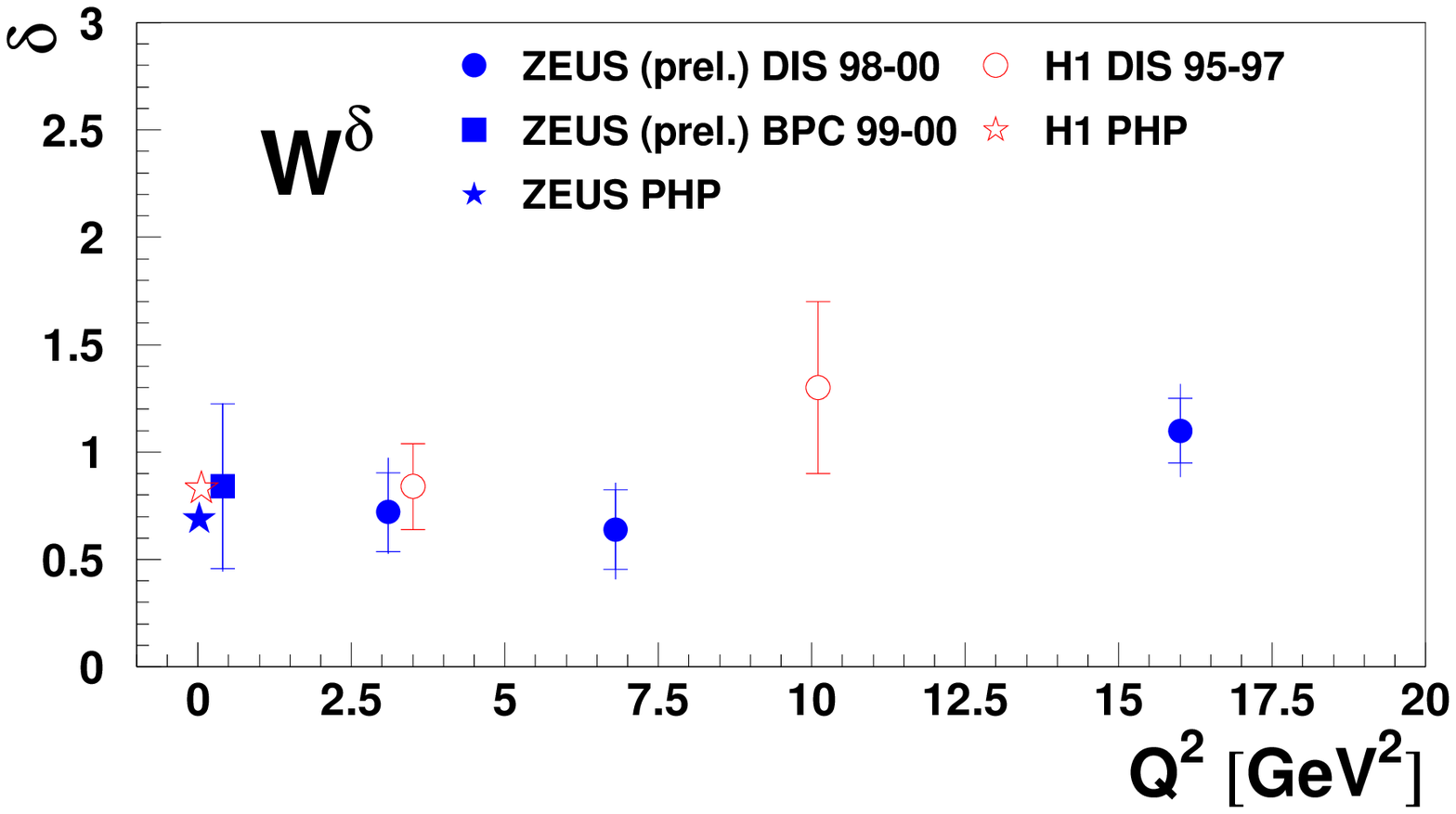}
\caption{The value of $\delta$ plotted as a function of $Q^2$
  obtained from fits to the form $\sigma_{J/\psi} \propto W^{\delta}$.}
\label{fig:disjpsi_delta.eps}
\end{minipage}
\end{figure}

\subsection{Test of SU(4) relation}
\begin{figure}[hbt]
\vspace*{-5mm}
\begin{minipage}{0.45\hsize}
\epsfxsize=1\hsize
\epsfbox{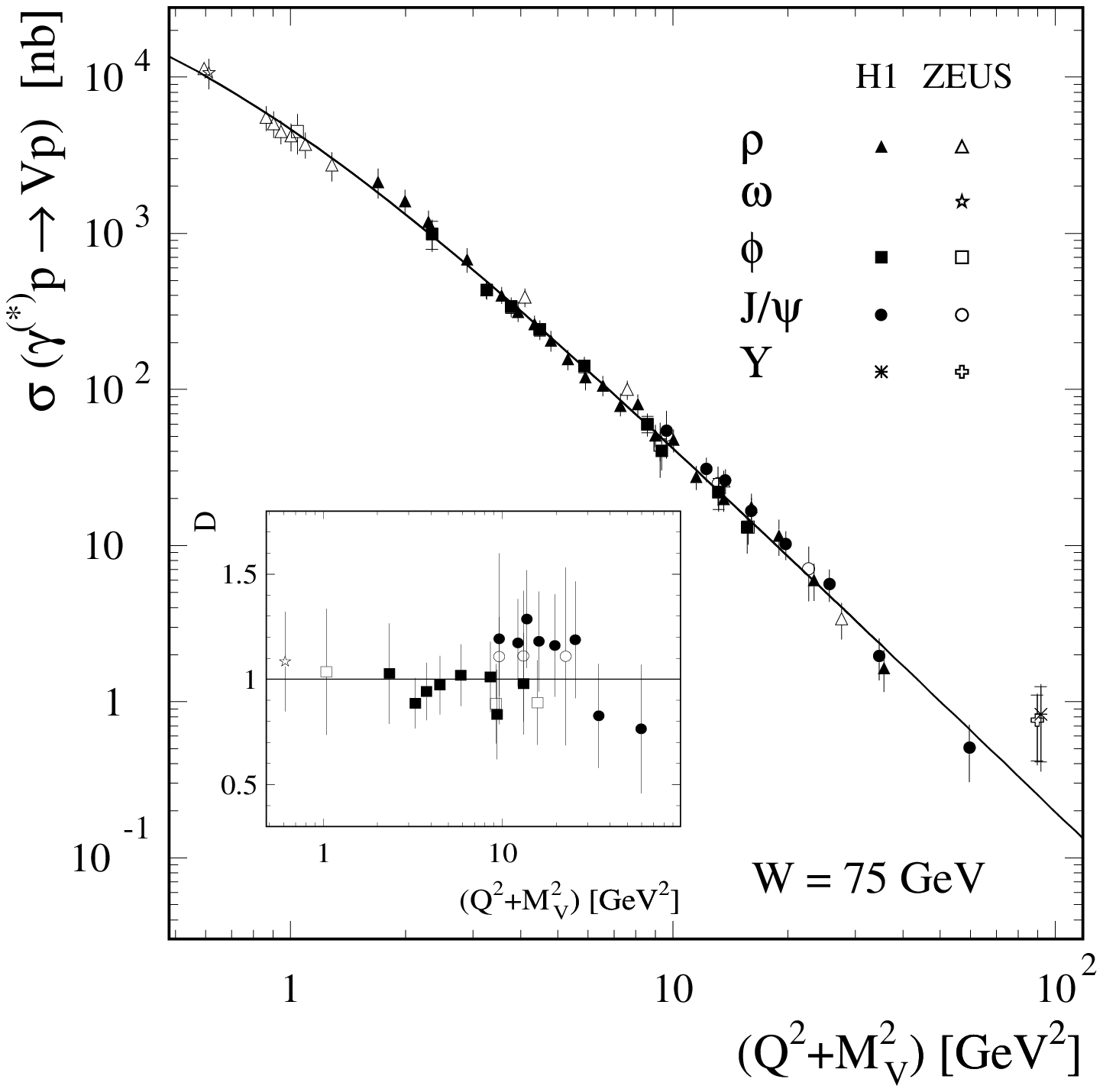}
\caption{The total cross sections for elastic V production scaled by 
  $SU(5)$ factors, as a function of ($Q^2+M^2_V$) at $W=75
  \units{GeV}$.  The curve corresponds to a fit to the H1 and ZEUS
  $\rho^0$ data. The insert shows the deviation of the parameterization
  from the data for $\omega$, $\phi$ and $J/\psi$.}
\label{fig:h100-070f3.eps}
\end{minipage}
\hspace*{2mm}
\begin{minipage}{0.45\hsize}
\vspace*{-6mm}
\epsfxsize=1\hsize
\epsfbox{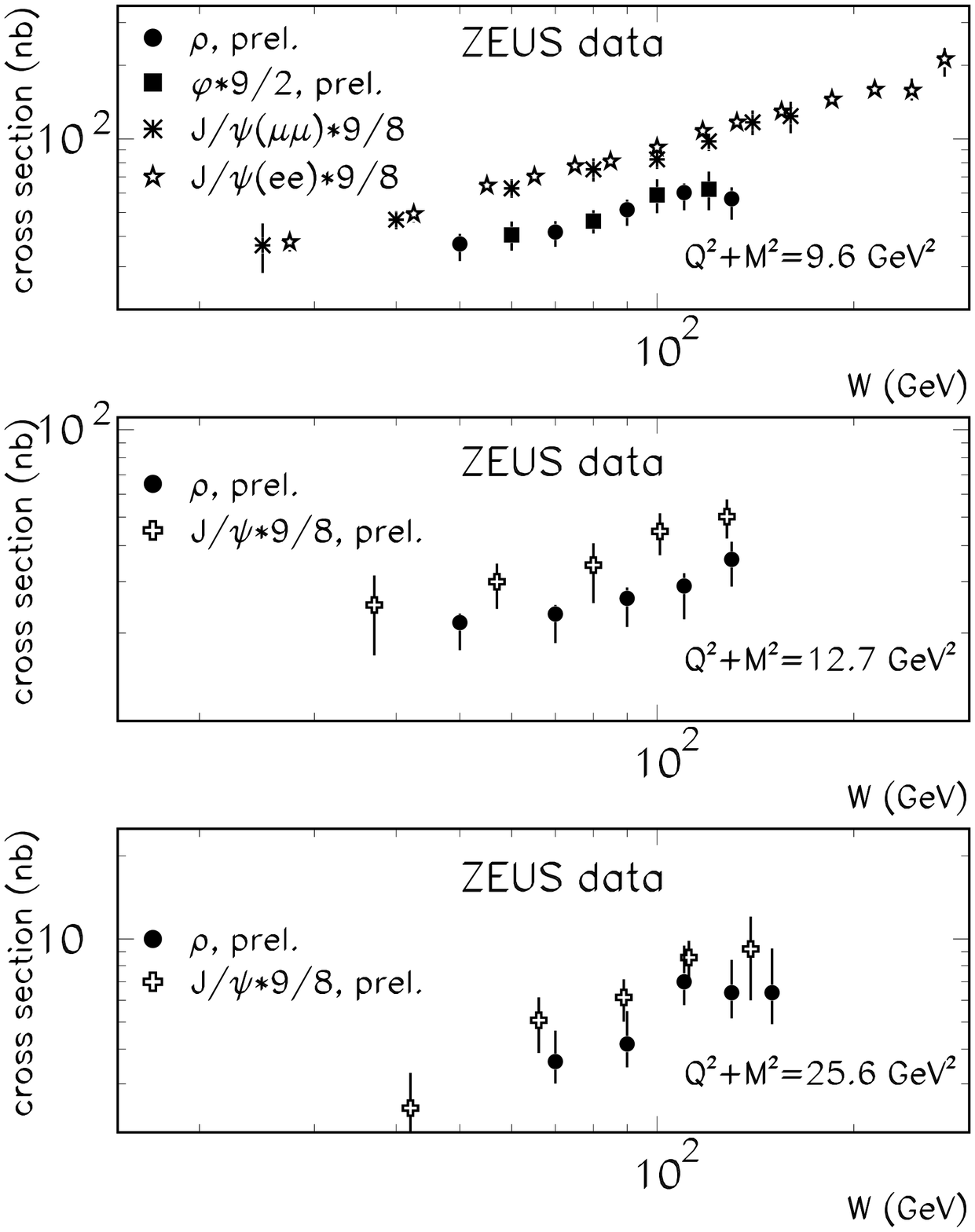}
\caption{Comparison of V weighted cross section values
at fixed $Q^2+M^2_V$ scales, as indicated in figure.}
\label{fig:vmheavy.eps}
\end{minipage}
\end{figure}

The H1 collaboration~\cite{h1apom} observed that when the cross
sections for various Vs, weighted by the appropriate $SU(4)$ factors
($\rho^0:\omega:\phi:J/\psi=9:1:2:8$) were plotted as function of
$Q^2+M_V^2$, at $W=75 \units{GeV}$, they seem to line up on a
universal curve, as shown in \fig{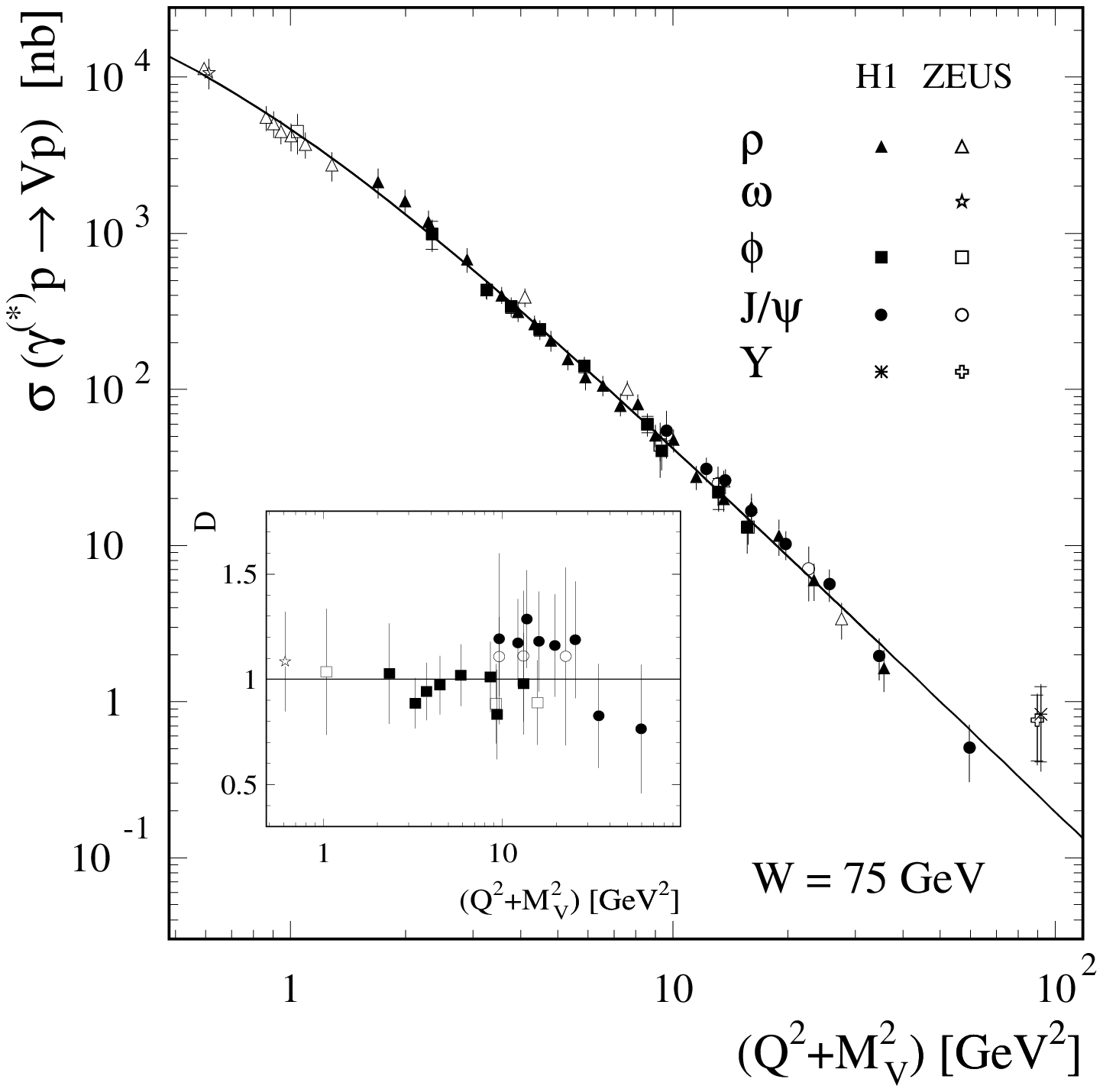}. However there was a
slight indication that the $J/\psi$ cross sections lie above the curve
(see the insert in \fig{h100-070f3.eps}). The new precise
measurements~\cite{ZEUS_J/psi_PHP,z-disjpsi} of $J/\psi$ production do
not fit this picture, as shown in \fig{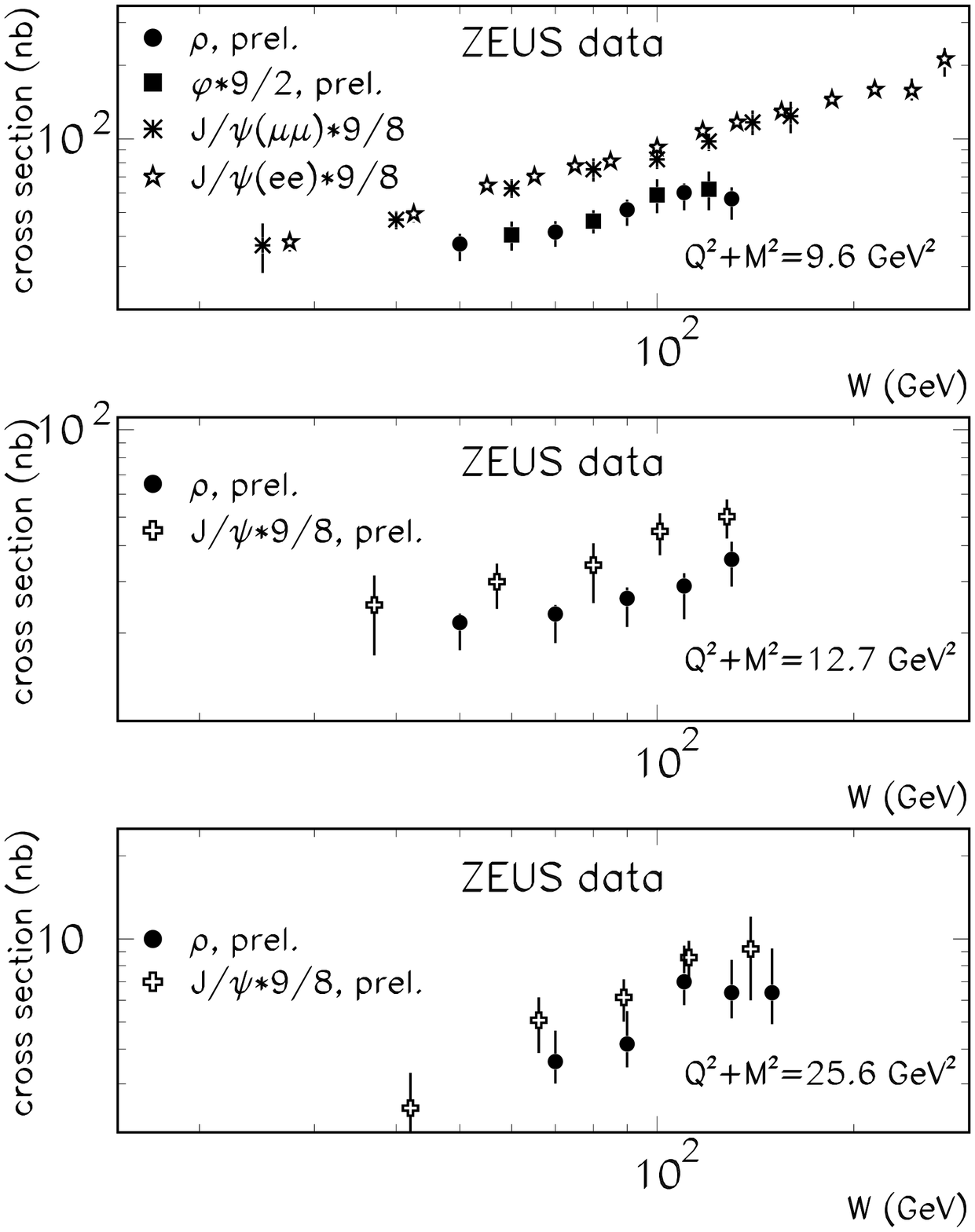}, where the $W$
dependence of light and heavy Vs, appropriately weighted by the
$SU(4)$ factor, is plotted in bins of $Q^2+M^2_V$. The $J/\psi$ data
lie far above the light Vs at all scales.  It seems, therefore, that
the variable $Q^2+M^2_V$ is a good variable for comparing the light Vs
but is not adequate when the $J/\psi$ is added to the comparison.

\subsection{Measurement of the Pomeron trajectory, 
$\alpha_{\pomsub}(t)$}
 
The {\LARGE $\pomsub$} trajectory, $\alpha_{\pomsub}(t)$, can be
measured by fitting the $W$ dependence of the $\gamma^\ast p$ cross
section in bins of $t$, $\frac{{\it d}\sigma}{{\it d}t} |_{t}(W) \sim
W^{\delta}$, where $\delta=4(\alpha_{\pomsub}(t)-1)$. The parameters
of the {\LARGE $\pomsub$} trajectory were extracted from measurements
of $\rho^0$~\cite{traj-rhopho,akdis01} and $J/\psi$ production
~\cite{h1apom,ZEUS_J/psi_PHP} and are summarized in
table~\ref{tab:pomtraj}. The measurements of $\alpha_{\pomsub}(t)$ and
the fitted trajectories are shown in \fig{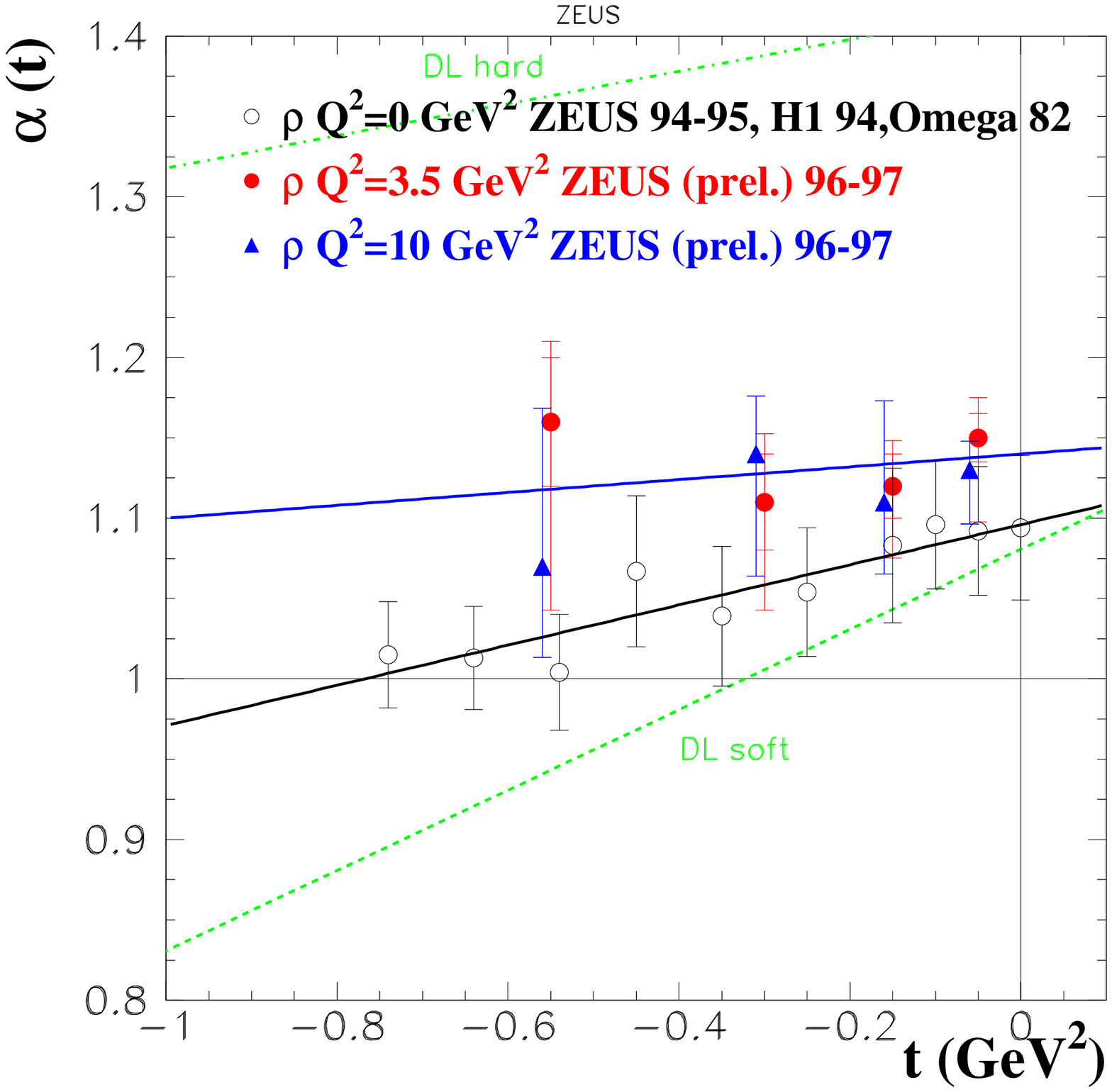}.  and in
\fig{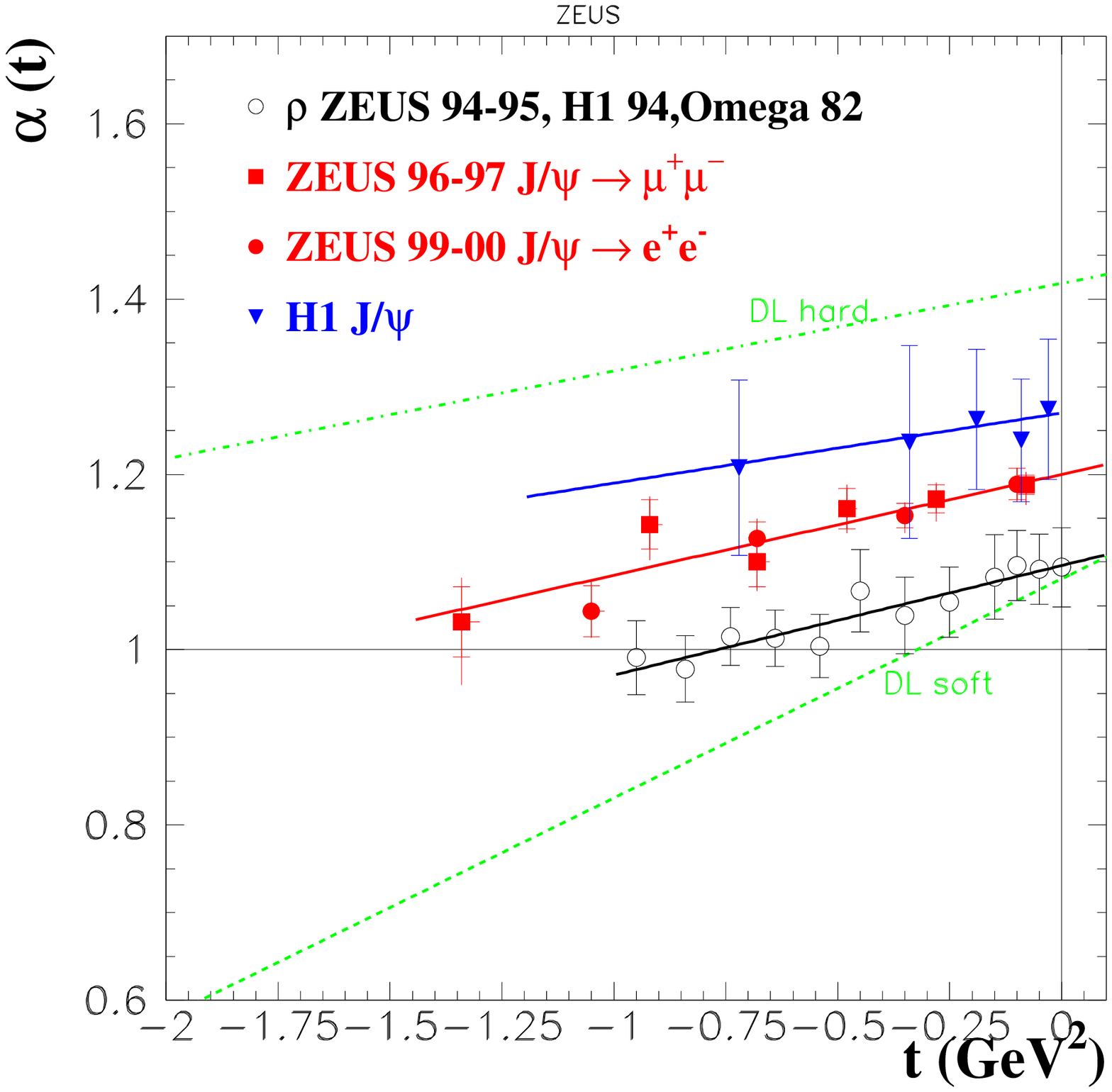}

The measured trajectories change with the $M_V^2$ and $Q^2$.

\begin{table}[hbt]
\caption{Compilation of  results obtained for $\apom(0)$ and $\aprime$.} 
\renewcommand\arraystretch{1.5}
\begin{tabular}{|l|r@{$\pm$}l|r@{$\pm$}l|}  \hline 
$V$ &\multicolumn{2}{c|}{$\apom(0)$} & \multicolumn{2}{c|} 
{$\aprime \units{(GeV^{-2})}$} \\ \hline
$\rho^0$, ($Q^2\simeq 0$) & 1.096 & 0.021 & 0.125 & 0.038  \\ \hline
$J/\psi$ (H1) & 1.27 & 0.05 & 0.08 & 0.017 \\ \hline
$J/\psi$ (ZEUS) & 1.200 & $0.009^{+0.004}_{-0.010}$ & 0.115 & $0.018^{+0.008}_{-0.015}$ \\ \hline
$\rho^0$ (ZEUS, DIS) & 1.14 & $0.01^{+0.03}_{-0.03}$ & 0.04 & $0.07^{+0.13}_{-0.04}$ \\ \hline
\end{tabular}
\label{tab:pomtraj}
\end{table}

\begin{figure}[htb]
\vspace*{-5mm}
\begin{minipage}{0.45\hsize}
\epsfxsize=1\hsize
\epsfbox{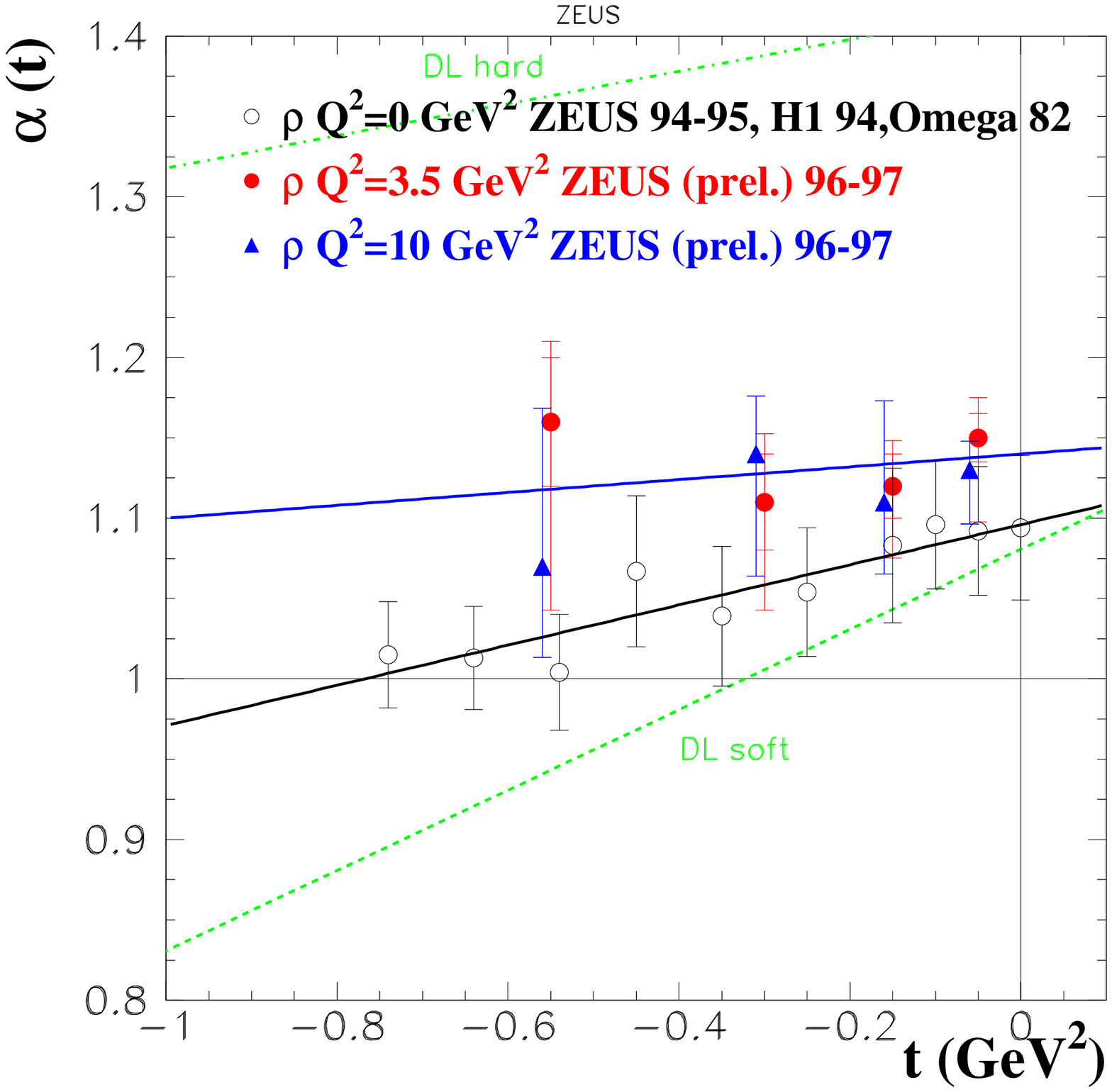}
\caption{
$\alpha_{\pomsub}(t)$, as measured for DIS $\rho^0$, for two $Q^2$
bins, $2<Q^2<6$ and $6<Q^2<40$ GeV$^2$, compared to $\rho^0$ 
photoproduction results. The solid lines are a fit to the data.}
\label{fig:gl_fit_syst2.eps}
\end{minipage}
\hspace*{2mm}
\begin{minipage}{0.45\hsize}
\epsfxsize=1\hsize
\epsfbox{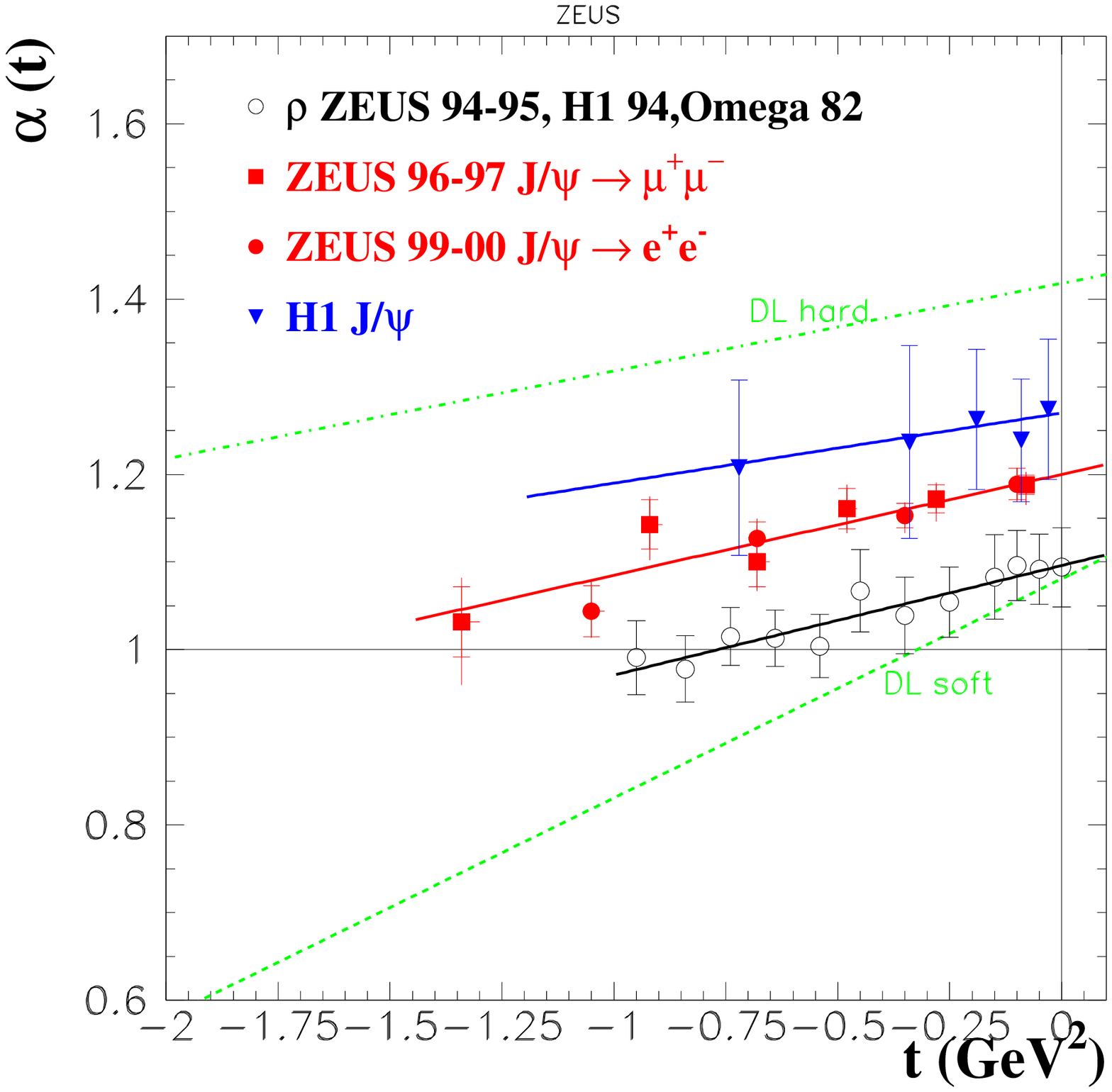}
\caption{The value of $\alpha_{\pomsub}(t)$ as a function of $t$ as
  measured for, $J/\psi$ photoproduction 
compared with previous
  results for $\rho^0$ photoproduction.} 
\label{fig:alpha_final.eps}
\end{minipage}
\end{figure}

\section{Angular distributions and helicity studies}

The separation  of the $\gamma^\ast p$ cross section into the
contribution from longitudinal photons,  $\sigma_L$, and 
transverse photons, 
$\sigma_T$, can be performed by measuring the V density
matrix elements, $r^{\alpha}_{ij}$~\cite{schilling}.  These matrix
elements are used to parameterize the angular distribution
$W(\cos{\theta_h},\phi_h,\Phi_h)$ of the helicity angles.
The density matrix elements, $r^{\alpha}_{ij}$, are related to 
the transition amplitudes of longitudinal or transverse photons
into longitudinal or transverse Vs. The transition amplitudes
are denoted as $T_{ij}$, where $i=0 (1)$ is a transverse 
(longitudinal) photon going into a $j=0 (1)$  transverse 
(longitudinal) V. The relations between the transition amplitudes
have been predicted by pQCD models~\cite{ivanov,royen,niko}.


\subsection{The ratio of longitudinal to transverse cross sections, 
  $R=\sigma_L/\sigma_T$}

The most extensive measurements~\cite{H199-010,akdis01} of $R(Q^2)$
were preformed for $\rho^0$ electroproduction.  The ratio $R$ of cross
sections induced by longitudinal polarized virtual photons to the
transversely polarized ones, $\sigma_L/\sigma_T$, is related to the
spin density matrix elements of $\rho^0$ through
\begin{displaymath}
R =\frac{1}
{\epsilon}\frac{r^{04}_{00}-\Delta^{2}}{1-(r^{04}_{00}-\Delta^{2})},
\end{displaymath}
where $r^{04}_{00}$ is a linear combination of the $\rho^0$ density
matrix elements, and $\Delta^{2}$ is proportional to the
contribution of the SCH non Conservation amplitude over the total
amplitude.  A small breaking of SCHC has been predicted~\cite{ivanov}
and measured~\cite{H199-010,zeusschc} ($\Delta=7.9 \pm 1.6
\units{\%}$~\cite{zeusschc}).

The $Q^2$ and $W$ dependence of $R$ is
plotted in \fig{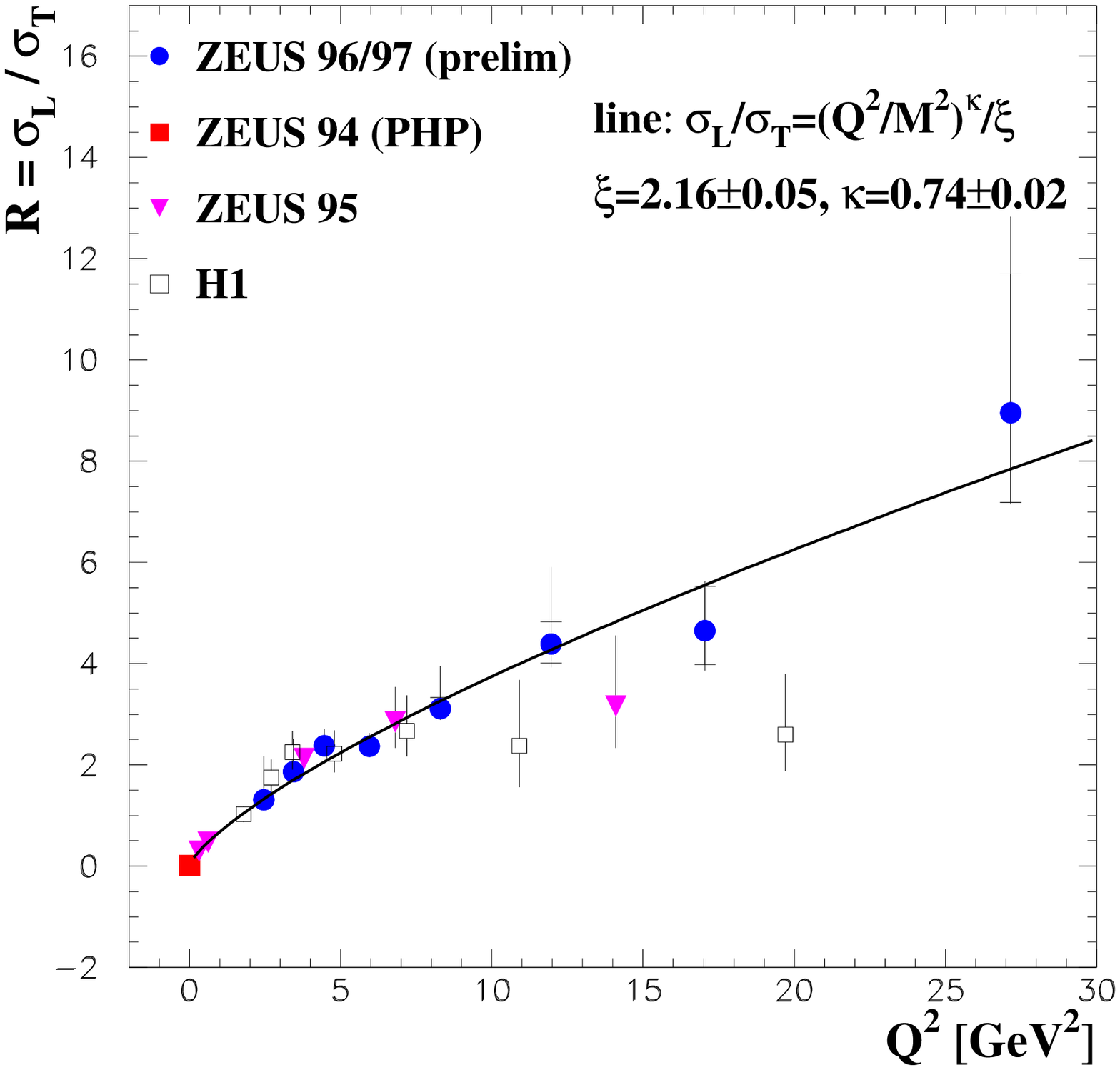} and \fig{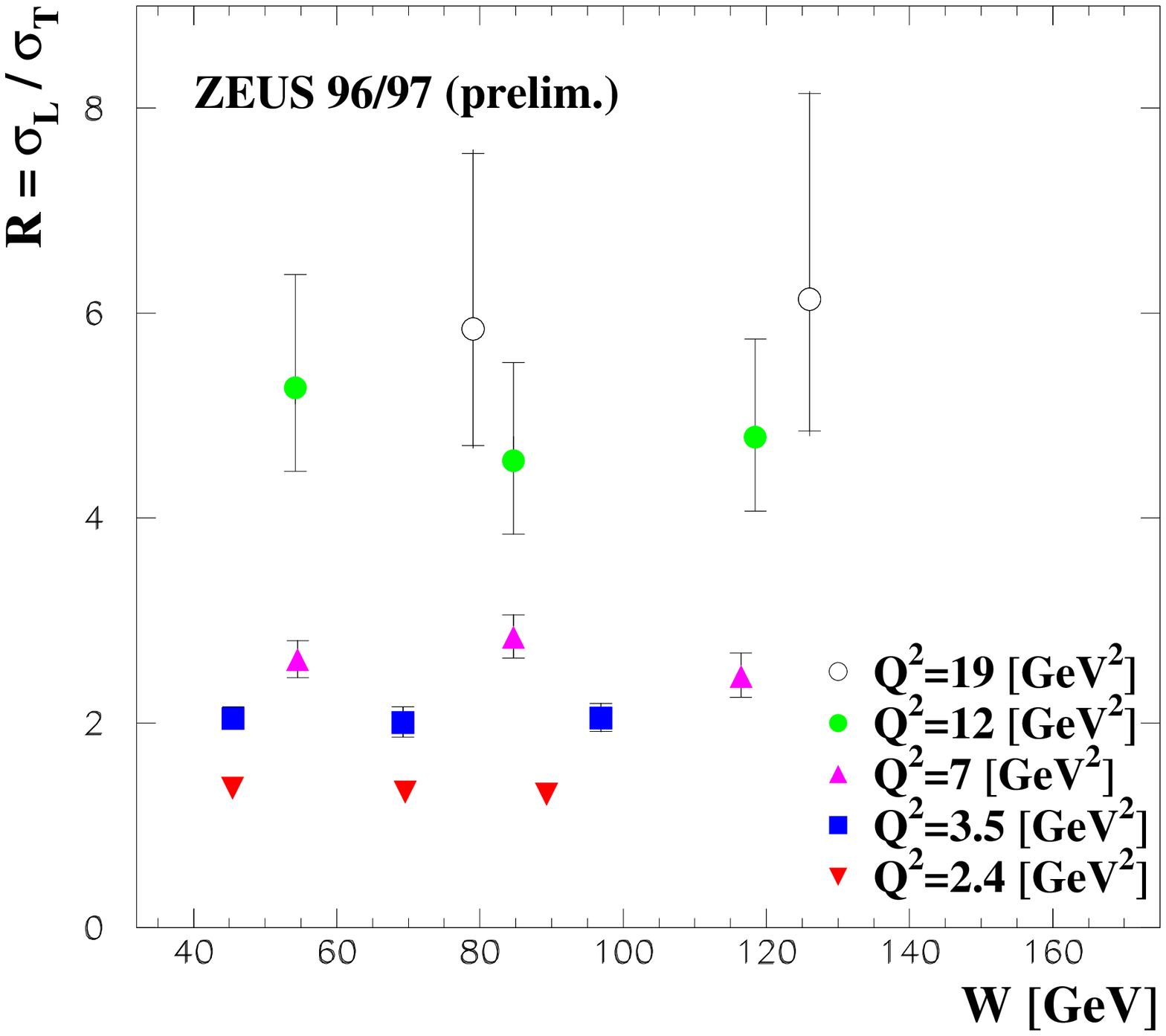}, respectively. 
A continuous rise of $R$ with
$Q^2$ is observed.  The $Q^2$ dependence of $R$ is well described
by a functional form
$R=\frac{1}{\xi}(Q^2/M_{\rho}^{2})^{\kappa}$ with $\xi=2.16\pm 0.05$ and
$\kappa=0.74\pm 0.02$~\cite{akdis01}.
 On the other hand, there seems to be no $W$
dependence of $R$ for fixed values of $Q^2$. This indicates that 
$\sigma_L$ and $\sigma_T$ have the same 
energy dependence, which is fixed by the $Q^2$ value
rather than the longitudinal or transverse configuration.

\begin{figure}[hbt]
\vspace*{-5mm}
\begin{minipage}{0.45\hsize}
\epsfxsize=1\hsize
\epsfbox{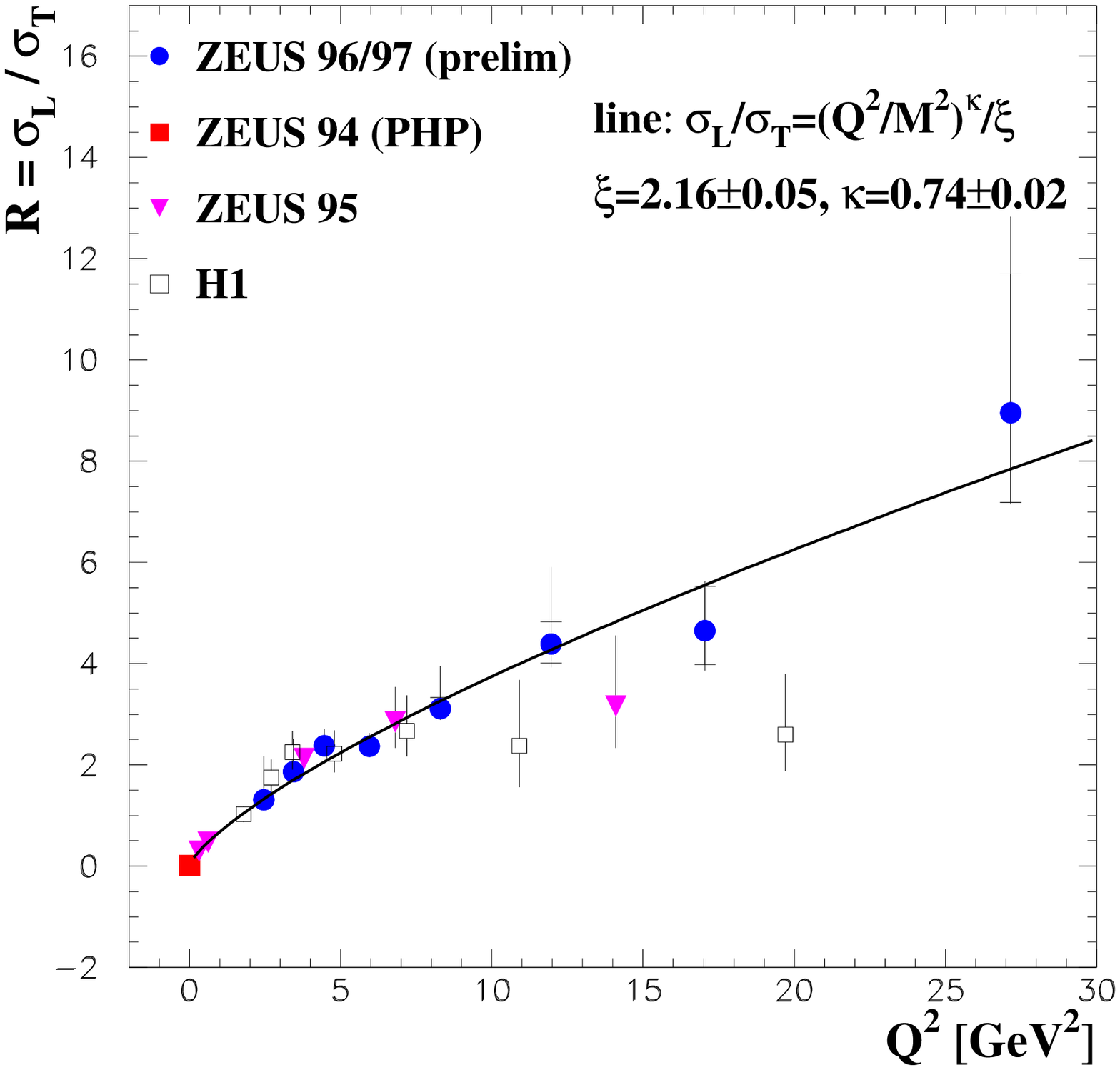}
\caption{The values of $R$ as a function of $Q^2$. }
\label{fig:R.eps}
\end{minipage}
\hspace*{2mm}
\begin{minipage}{0.45\hsize}
\vspace*{-6mm}
\epsfxsize=1\hsize
\epsfbox{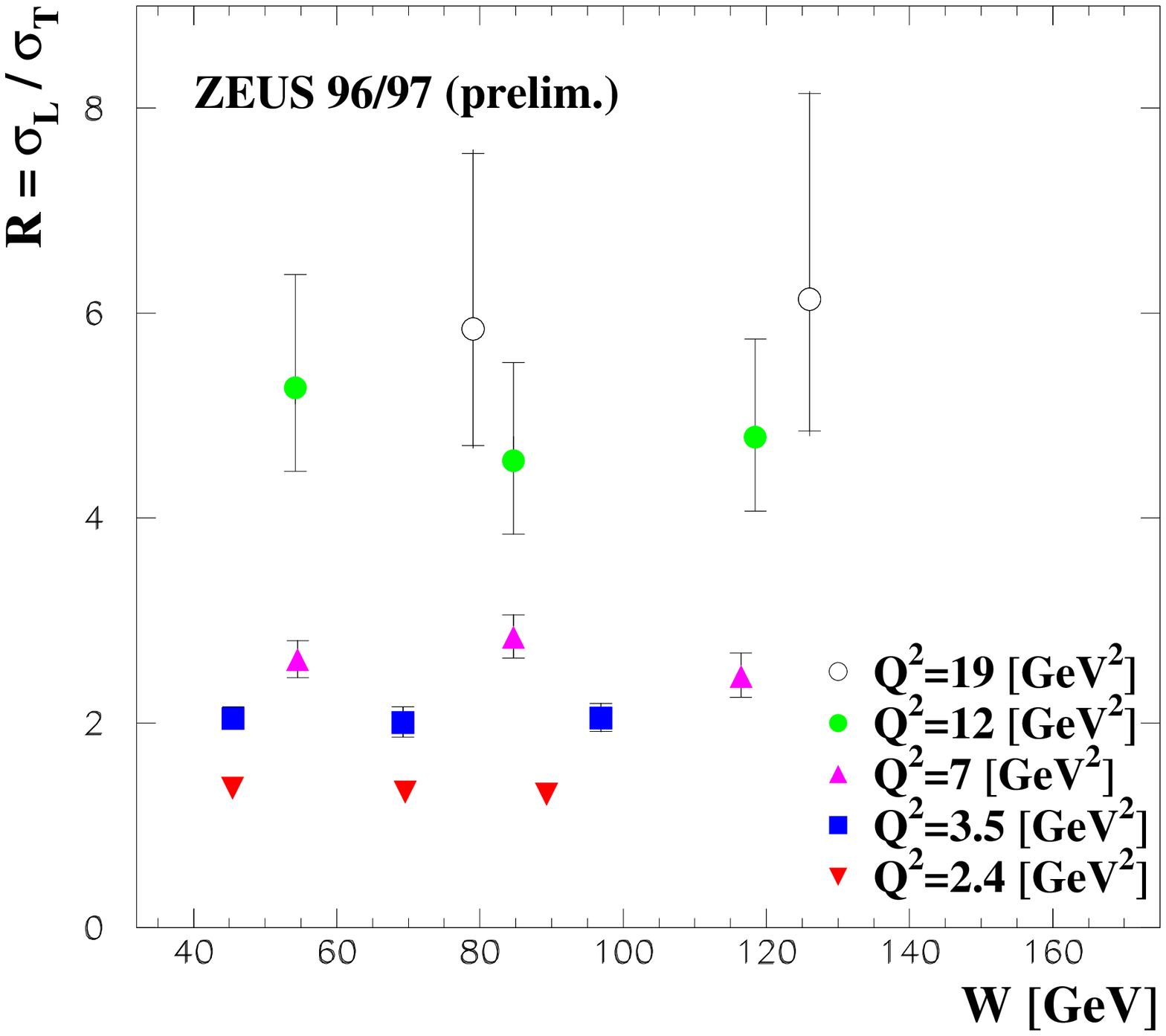}
\caption{The value of $R$
  is presented as a function of $W$, for different $Q^2$ values. }
\label{fig:wR.eps}
\end{minipage}
\end{figure}

\subsection{A measurement of the $t$ dependence of the helicity structure}

The extracted~\cite{h1_t_helicity} values of \rfour\ for $\rho^0$
production are presented in \fig{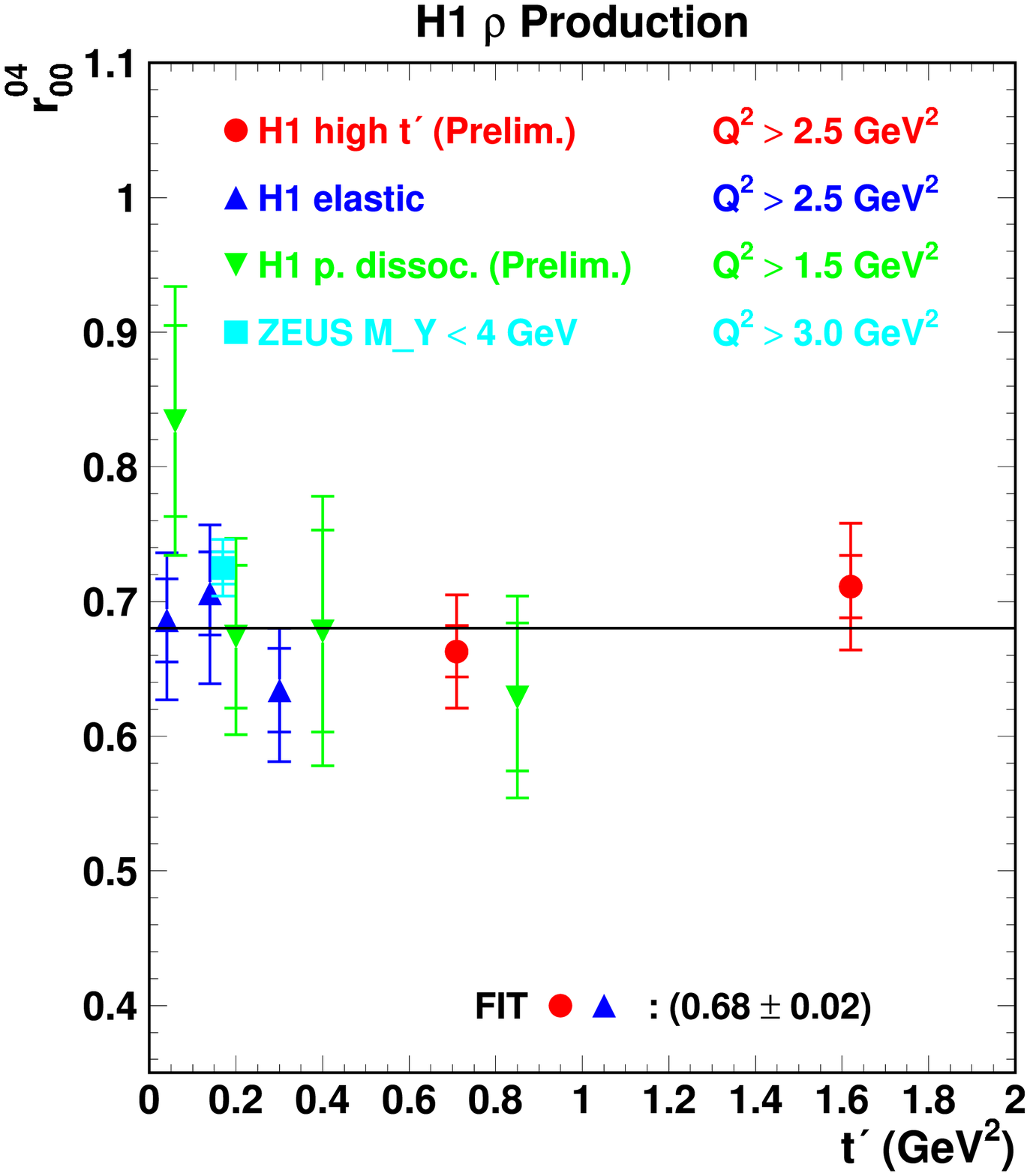}, where $r^{04}_{00}
\propto \frac{|T_{00}|^2+|T_{01}|^2}{N} \propto
\sigma_{L}/(\sigma_{T}+\sigma_{L})$ and $N=\sum_{i,j}|T_{ij}|^2$.  No
significant variation of \rfour\ with $\tprim=|t-t_{min}|$ is observed.
This observation implies that the slopes
of the exponentially falling $t$ distributions for the transverse and
longitudinal $s$-channel helicity conserving amplitudes, $T_{00}$ and
$T_{11}$, are very similar.

\begin{figure}[htb]
\vspace*{-5mm}
\begin{minipage}{0.45\hsize}
\epsfxsize=1\hsize
\epsfbox{h1rho_r04t.eps}
\caption{Measurement of \protect\rfour\
 as a function of
\protect\tprim. 
The  curve is a result of fitting the data with calculation 
of~\protect\cite{ivanov}.}
\label{fig:h1rho_r04t.eps}
\end{minipage}
\hspace*{2mm}
\begin{minipage}{0.45\hsize}
\vspace*{-6mm}
\epsfxsize=1\hsize
\epsfbox{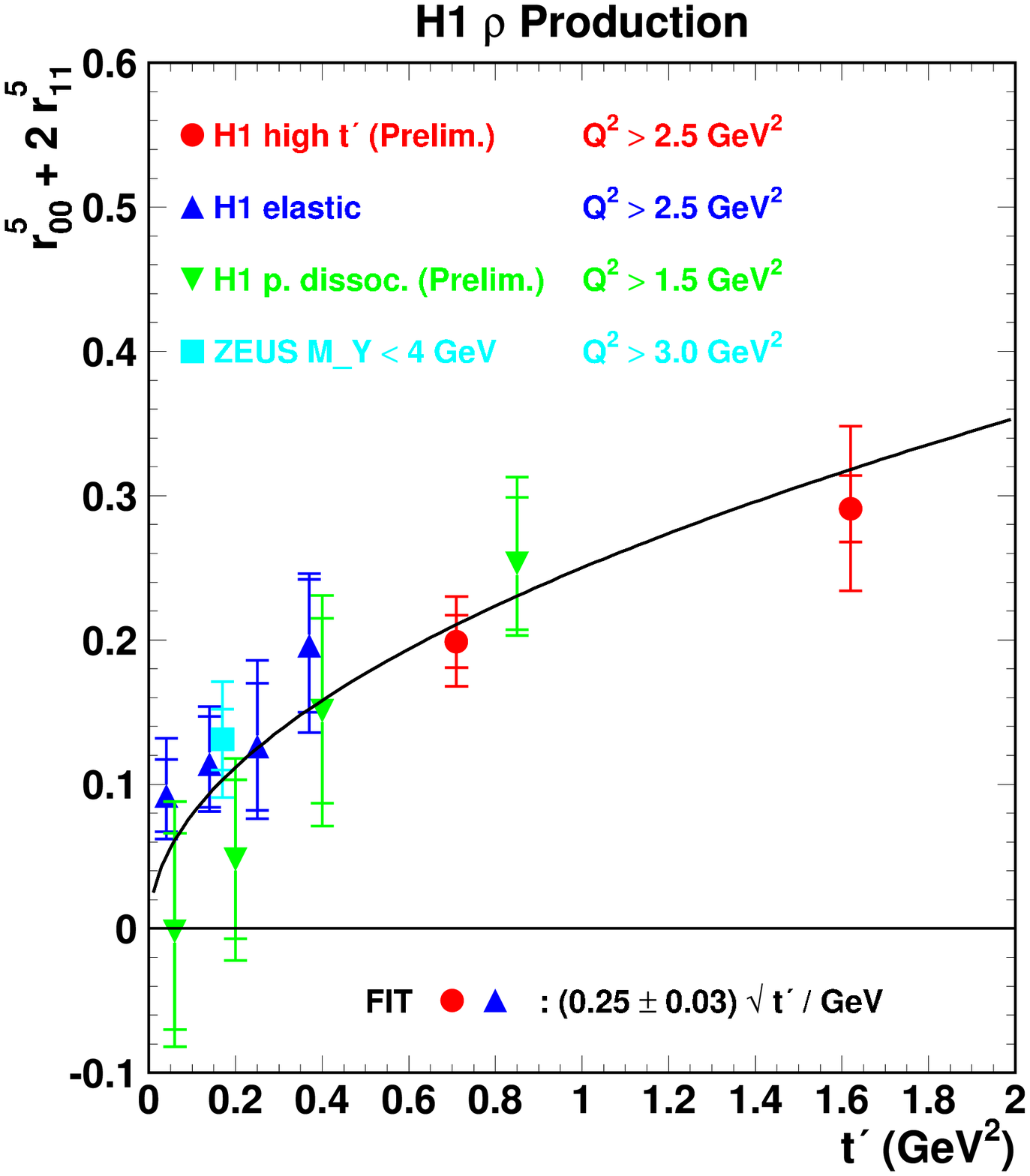}
\caption{Measurement of \protect\rfivecomb\ as a function of
\protect\tprim.
The  curve is a result of fitting the data with calculation 
of~\protect\cite{ivanov}. For SCHC a null result is expected, 
independent of \protect\tprim. }
\label{fig:h1rho_r5t.eps}
\end{minipage}
\end{figure}

The combination $r^{5}_{00}+2r^{5}_{11}$ is presented in
\fig{h1rho_r5t.eps}~\cite{h1_t_helicity}.  $r^{5}_{00}$ is
expected~\cite{ivanov,royen,niko} to be proportional to the product of
the dominant non-flip amplitude $T_{00}$ and the single flip $T_{01}$
amplitude, $r^{5}_{00}\propto \frac{1}{N} Re( T_{00}T_{01})$. The
$T_{01}$ amplitude is expected to be the largest helicity flip
amplitude. The $r^5_{11}$ matrix element, on the other hand is
proportional to the non-dominant amplitudes $r^{5}_{11} \propto
\frac{1}{N}Re(2T_{11}T_{10}^{\dagger}-2T_{10}T_{1-1}^{\dagger})$.  The
strong \tprim\ dependence of the \rfivecomb\ combination is thus
attributed mainly to the predicted~\cite{ivanov,royen,niko}
$\sqrt{\tprim}$ dependence of the ratio of  $T_{01}$ to the
non-flip amplitudes.

The values for \ronecomb\ are shown in \fig{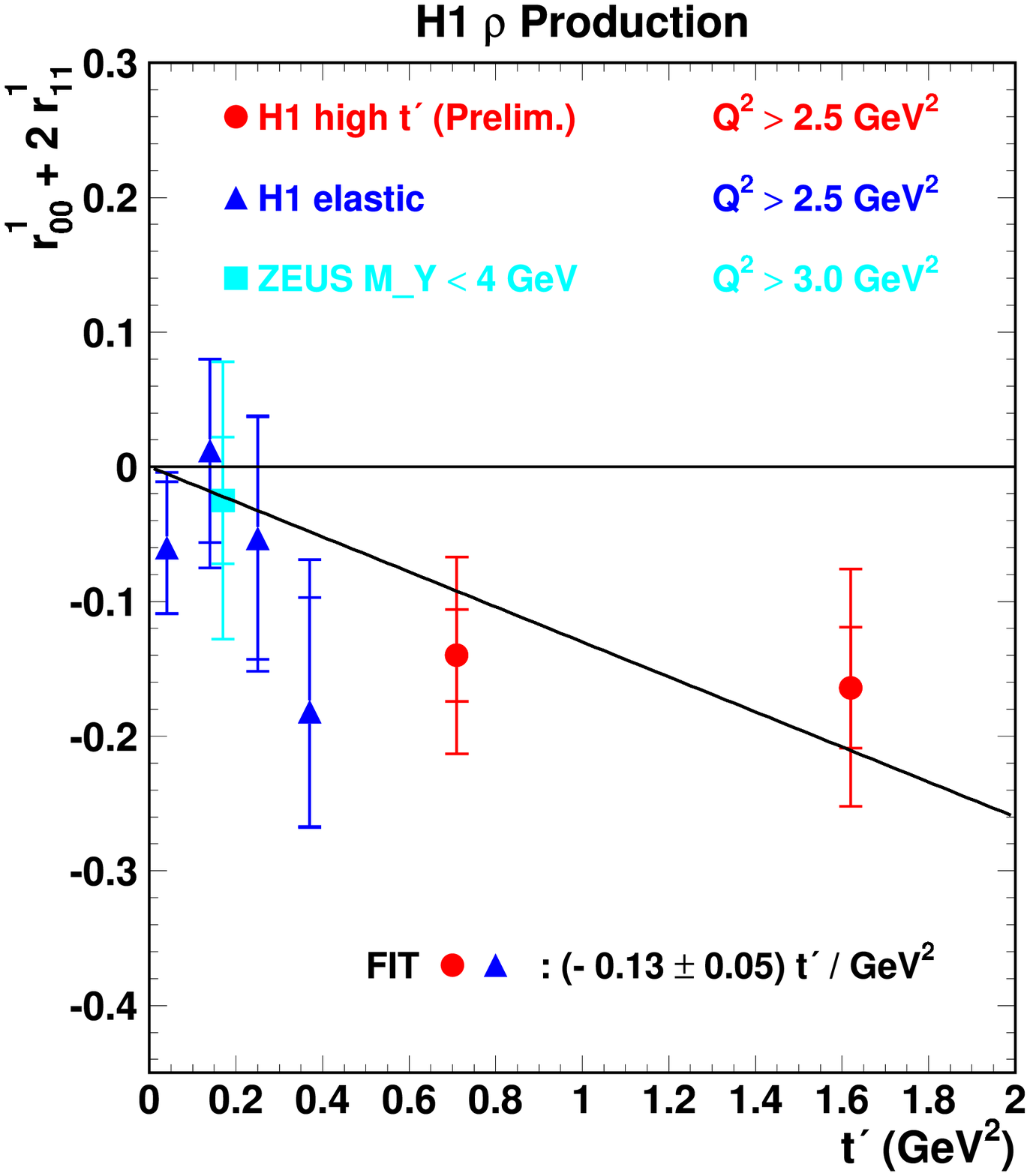}. The matrix
element $r^{1}_{00}$ is proportional to the single flip amplitude
$T_{01}$ square, $r^{1}_{00} \propto -\frac{|T_{01}|^2}{N}
\propto(\sqrt{t})^2$, while the matrix element $r^{1}_{11}$ is
expected to have the same $t$ dependence as the double flip amplitude
$T_{1-1}$, $r^{1}_{11} \propto \frac{1}{N}(T_{11}T_{1-1}^{\dagger})
\propto t$, therefore the \tprim\ dependence of the combination
\ronecomb\ is expected to be linear, up to effects of the single and
double-flip amplitudes in the denominator $N$.  The combination
\ronecomb\ is significantly different from zero and negative, which
implies violation of SCHC.  The sign of the combination gives
information on the relative strength of the $T_{01}T_{01}^\dagger$ and
$T_{11}T_{1-1}^\dagger$ products of amplitudes,
$|T_{01}|^2>(T_{11}T_{1-1}^{\dagger})$, and therefore
$T_{01}>T_{1-1}$.  It confirms that the $T_{01}$ amplitude is
significantly larger than the double flip amplitude in the present
kinematic domain.

\begin{figure}[hbt]
\begin{center}
\epsfxsize=0.7\hsize
\epsfbox{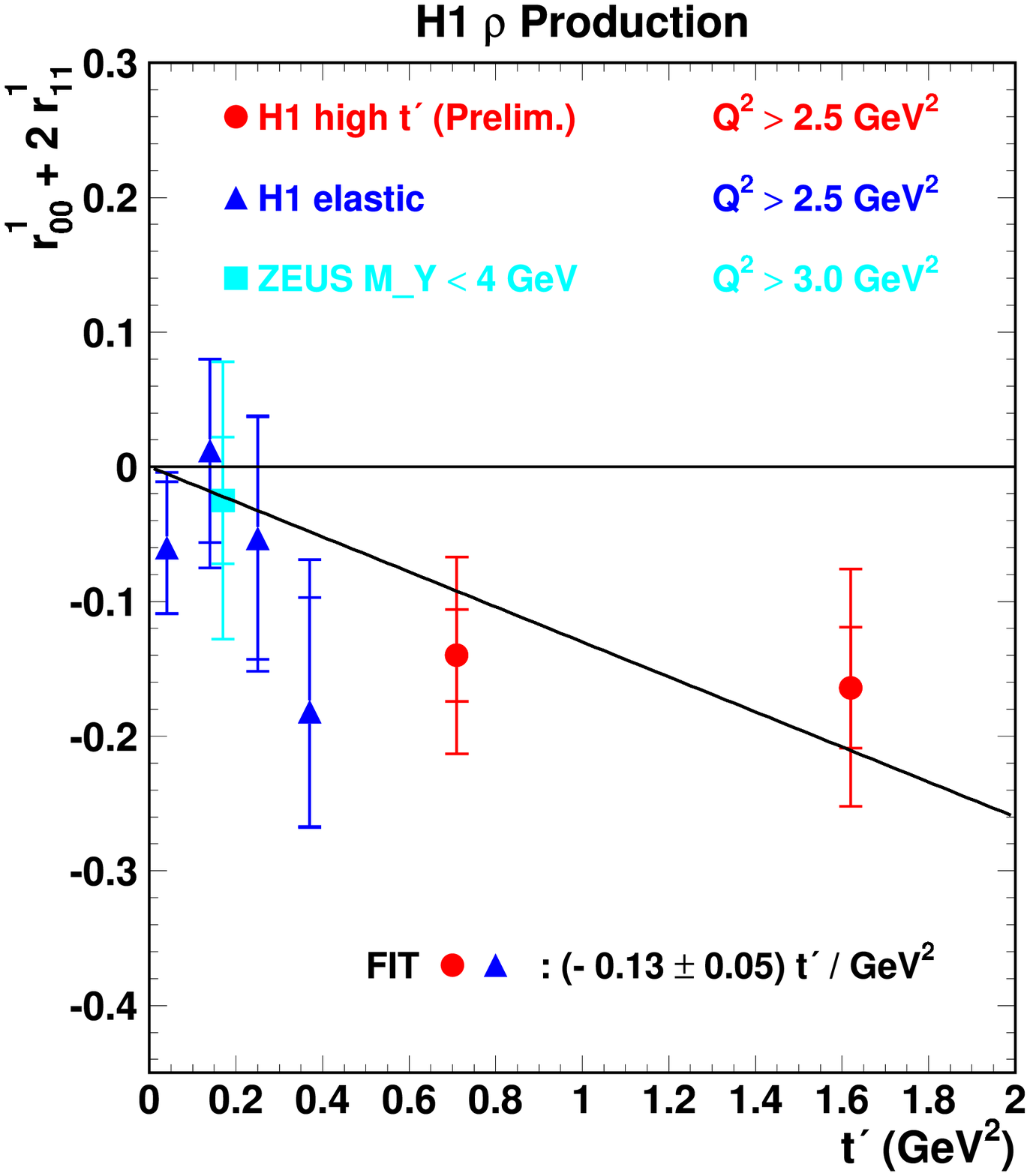}
\caption{Measurement of \protect\ronecomb\ as a function of
\protect\tprim.
The  curve is a result of fitting the data with calculation 
of~\protect\cite{ivanov}. For SCHC a null result is expected, 
independent of \protect\tprim. }
\label{fig:h1rho_r1t.eps}
\end{center}
\end{figure}

\section{ the differential cross section at low \tlc\ }

The differential cross section as function of $t$ has been studied for
$\rho^0$ electroproduction.  The values of $b(Q^2)$ are extracted from
a fit to an exponential form, $\frac{{\it d}\sigma}{{\it d}t} \propto
\exp{(bt)}$, for $t<1 \units{GeV^2}$ in elastic production $(b_{el})$
and for $t<2 \units{GeV^2}$ in proton dissociation $\rho^0$ production
$(b_{pd})$, for different $Q^2$ regions .  The data cover a kinematic
range of $50<W<140 \units{GeV}$ and $2<Q^2<80 \units{GeV^2} $
(elastic) and $2<Q^2<50\units{GeV^2}$ (proton dissociation).

\begin{figure}[hbt]
\vspace*{-5mm}
\begin{minipage}{0.45\hsize}
\epsfxsize=1\hsize
\epsfbox{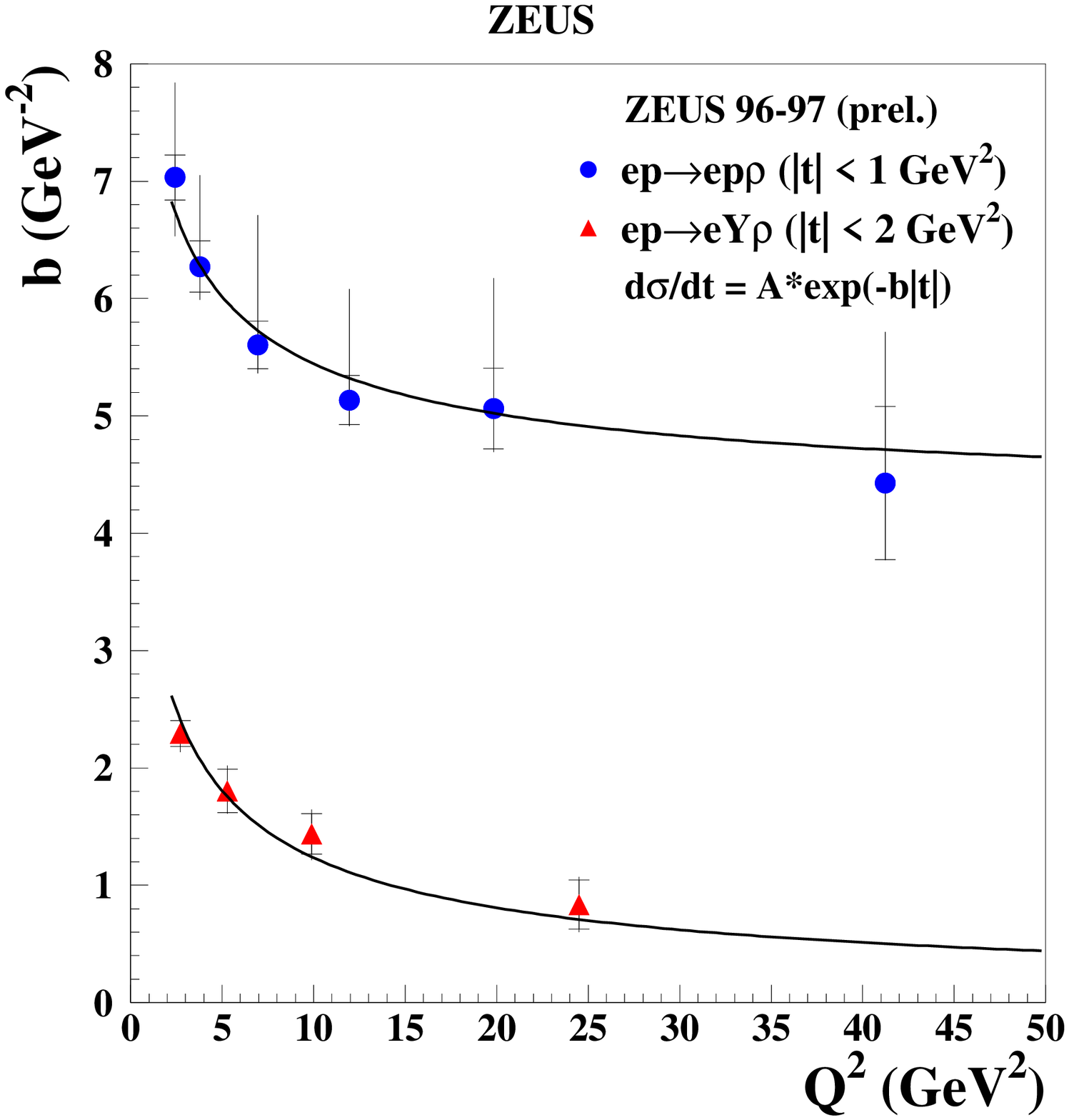}
\caption{The slope $b$ of exclusive and proton-dissociative 
electroproduction of $\rho^0$.}
\label{fig:pdis_el_bq2.eps}
\end{minipage}
\hspace*{2mm}
\begin{minipage}{0.45\hsize}
\vspace*{-6mm}
\epsfxsize=1\hsize
\epsfbox{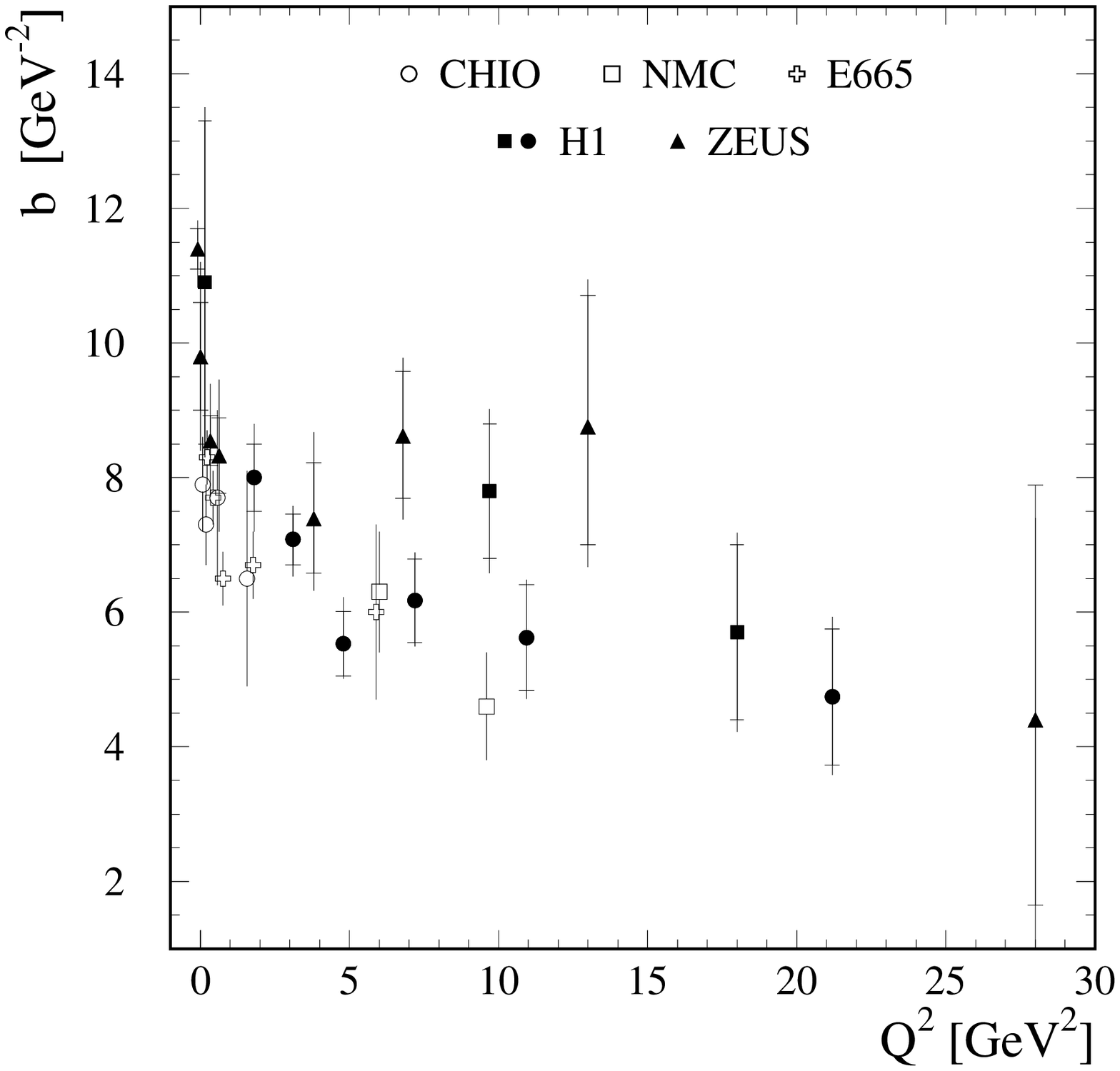}
\caption{Exponential slope $b$ of the
$t$ dependence for elastic $\rho^0$ production
as a function of $Q^2$.}
\label{fig:h199-010f19.eps}
\end{minipage}
\end{figure}

The values of $b_{el}$ and $b_{pd}$ are shown  in \fig{pdis_el_bq2.eps}
as a function of $Q^2$.
The $Q^2$ dependence of $b$ is well described by the following
functional forms,
\begin{eqnarray*}
b_{pd}(Q^2) & = & A/(M_V^2(1+R(Q^2))) \, ,\\  
b_{el}(Q^2) & = & b_{pd}(Q^2)+b_{\infty} \, ,\\
R=\frac{\sigma_L}{\sigma_T} & = & \frac{1}{\xi}(Q^2/M_V^2)^\kappa \, ,\\
A & = & 3.46 \pm 0.14 \, , \\   
b_{\infty} & = & 4.21 \pm 0.12 \units{GeV^{-2}} \, .
\end{eqnarray*}
One clearly sees the decrease in the value of $b_{el}$ with $Q^2$,
approaching an asymptotic value, from fit $b_{\infty}=4.21 \pm 0.12
\units{GeV^{-2}}$. The slopes $b_{el}$ and $b_{pd}$ are described with
the same $Q^2$ dependence, in compliance with the expectations of
vertex factorization~\cite{factorization}. It is interesting to note
that the ad hoc parametrization of $b_{pd}$, based on the $Q^2$
dependence of $R$, describes well the measurements of $b_{pd}$ and
$b_{el}$. The
parameterization is motivated by the expected $Q^2$ dependence of the
cross section and the relation between the probability for
longitudinal configuration and the size of the $q\bar{q}$ pair.
   
The measurements of $b_{el}$, done  by H1~\cite{H199-010} 
(\fig{h199-010f19.eps}), show 
the same  $Q^2$ behavior. It is also evident that at low $Q^2$, the HERA
measurements lie systematically above the low energy
fixed target results. This might indicate shrinkage of the diffractive
peak, at low $Q^2$, as $W$ increases.

In order to test the factorization hypothesis at the proton
vertex~\cite{factorization}, the ratio of ${\it d}\sigma/{\it d}t$ of
the elastic and proton dissociation reactions is studied as function
of $Q^2$, for two fixed values of $t$, as shown in
\fig{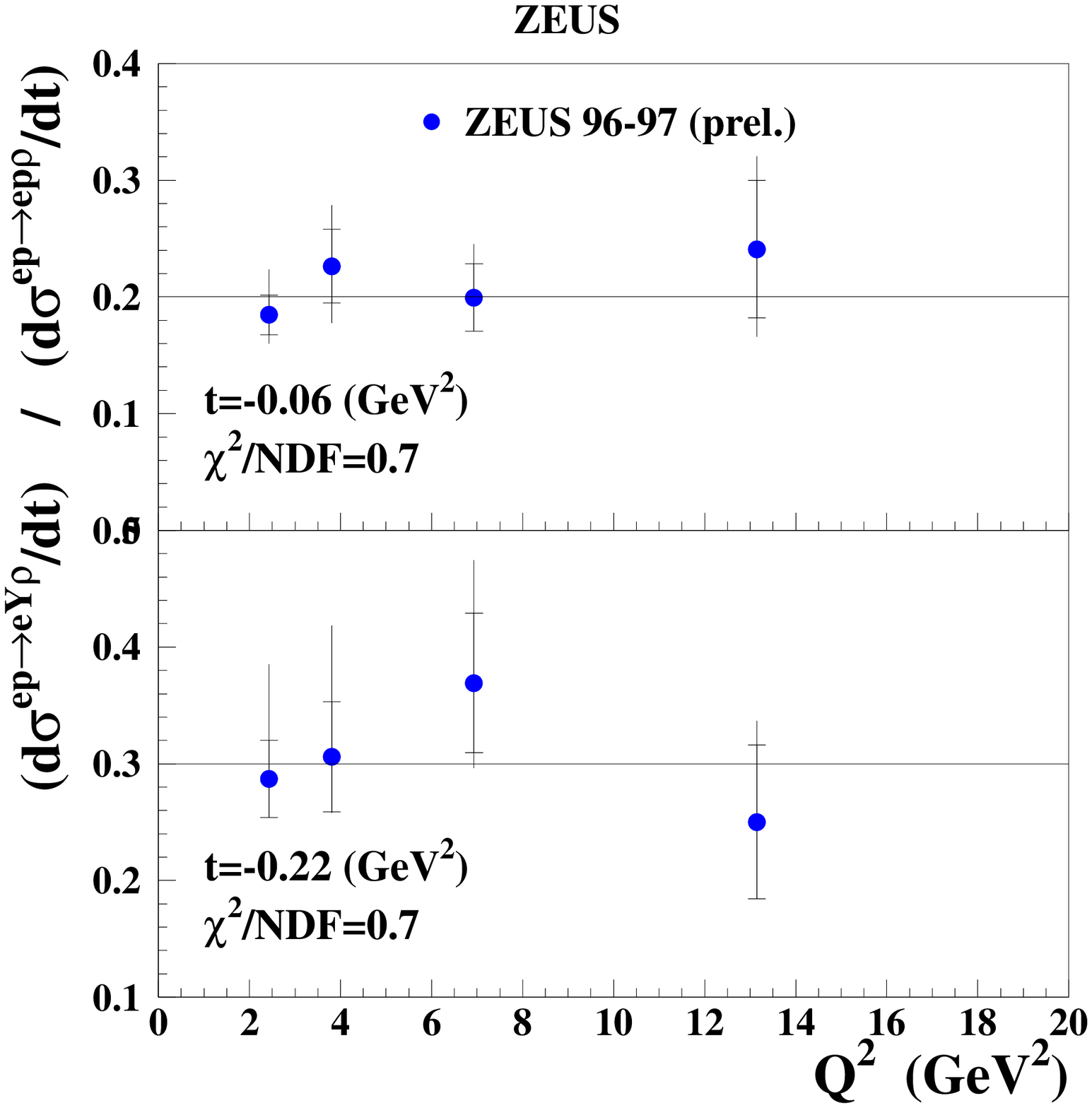}.  The ratio is consistent with being $Q^2$
independent. This lends support to the hypothesis that factorization
holds at the proton vertex for $\rho^0$ electroproduction in the
region $2<Q^2<20 \units{GeV^2}$ and $|t|<0.4 \units{GeV^2}$.
        
\begin{figure}[hbt]
\begin{center}
\epsfxsize=0.5\hsize
\epsfbox{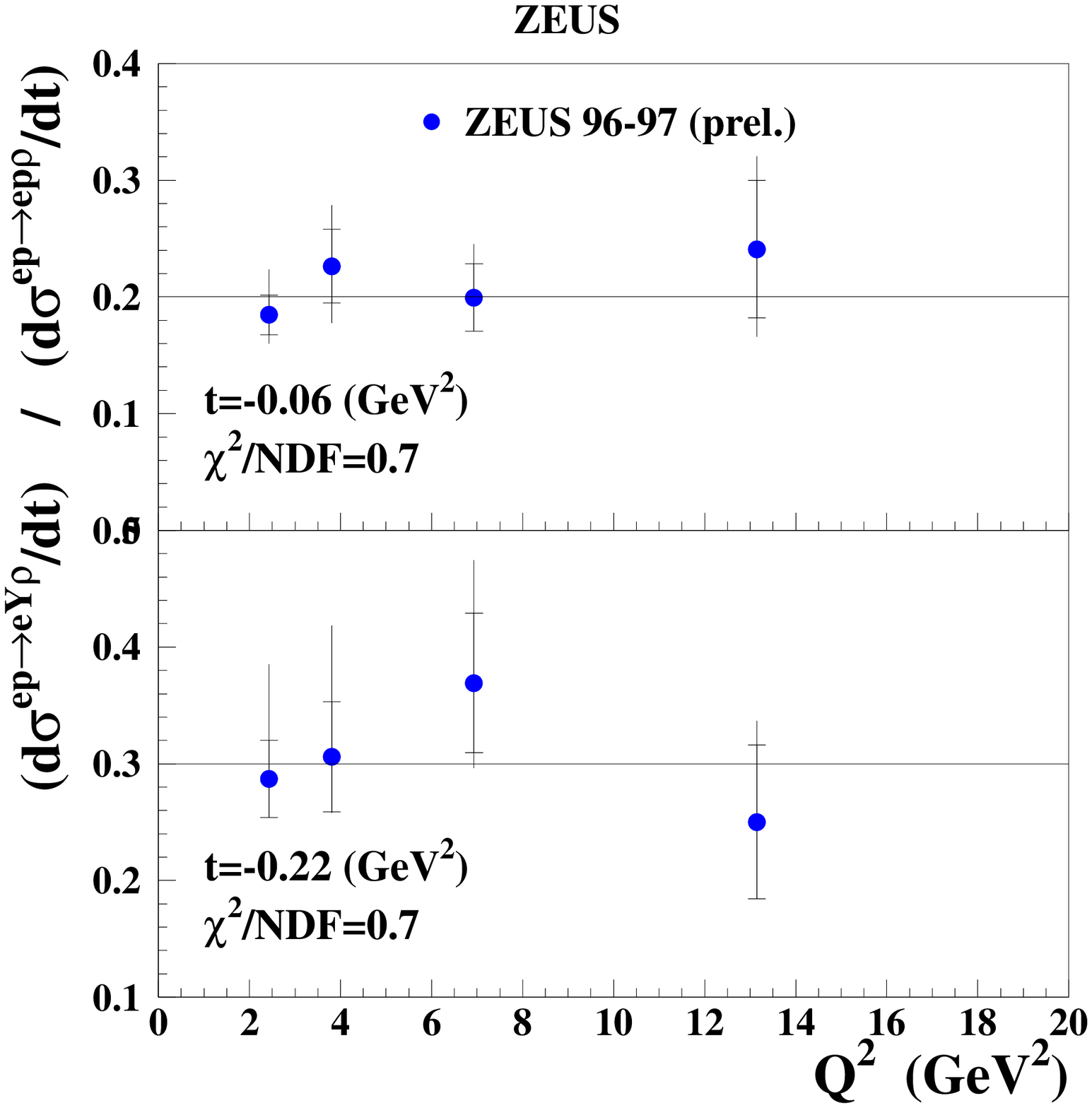}
\caption{Ratio of proton-dissociative and exclusive $\rho^0$ 
  electroproduction processes as function of $Q^2$ for fixed $t$
  values, as given in the figure. The lines are the results of a best
  fit to a constant ratio at each $t$ value.}
\label{fig:pdis_el_rat.eps}
\end{center}
\end{figure}

\section{Conclusions} 

It is evident that the hardness of the process in the elastic
production of vector mesons in $ep$ collisions at high energy, is
influenced by both initial (e.g. $Q^2$) and final state configurations
(e.g. $M_V^2$).  This is demonstrated by the change of the energy
dependence with $Q^2$ and $M_V^2$, and by the decrease of the $b$
slope with increasing $Q^2$ and $M_V^2$. Longitudinal and transverse
photons seem to have the same energy dependence at fixed $Q^2$, as
well as the same $t$ dependence.  This indicates that the $q \bar{q}$
pairs initiated by a longitudinal or transverse photon, and
contributing to the production of a given vector meson, are typically
of the same size.  This size is determined both by the virtualty $Q^2$
of the fluctuating photon and by the  probability to form the
 vector meson final state.

Although the overall picture is consistent with pQCD predictions, 
the quantitative calculations need further improvements.

\section{Acknowledgments}
I am very grateful to Alberto Santoro and the LISHEP organizing committee
for their financial support and hospitality 
during the LISHEP advanced school on HEP and the following workshop
on diffractive physics.

This research was supported in part by the Israel Science Foundation
(ISF), by the German Israel Foundation (GIF) and by Israel Ministry
of Science.

\newpage

\end{document}